\documentclass[iop,twocolumn,natbib209]{emulateapj}  
\usepackage{graphicx}
\usepackage{natbib}
\usepackage{lscape} 
\usepackage{hyperref}
\usepackage{CJK}

\newcommand{\myemail}{ywang@pmo.ac.cn}

\shorttitle{Different evolutionary Stages in the W3 Main Complex}
\shortauthors{Wang et al.}

\begin{document}
\begin{CJK*}{UTF8}{gbsn}


\title{Different Evolutionary Stages in the Massive Star Forming Region W3 Main Complex}


\author{Yuan Wang (王渊)\altaffilmark{1, 2, 3}, Henrik Beuther\altaffilmark{2}, Qizhou Zhang\altaffilmark{4}, Arjan Bik\altaffilmark{2}, Javier A. Rod\'on\altaffilmark{5}, Zhibo~Jiang (江治波)\altaffilmark{1}, Cassandra~Fallscheer\altaffilmark{6,7}}

\altaffiltext{1}{Purple Mountain Observatory \& Key Laboratory for Radio Astronomy, Chinese Academy of Sciences, Nanjing 210008, China, \myemail}
\altaffiltext{2}{Max-Plank-Institute for Astronomy, K\"onigstuhl 17, Heidelberg 69117, Germany}
\altaffiltext{3}{Graduate University of the Chinese Academy of Sciences, 19A Yuquan Road, Shijingshan District, Beijing 100049, China}
\altaffiltext{4}{Harvard-Smithsonian Center for Astrophysics, 60 Garden Street, Cambridge, MA 02138, USA}
\altaffiltext{5}{European Southern Observatory, Alonso de C\'ordova 3107, Vitacura, Casilla 19001, Santiago 19, Chile}
\altaffiltext{6}{Department of Physics and Astronomy, University of Victoria, PO Box 355, STN CSC, Victoria BC, V8W 3P6, Canada}
\altaffiltext{7}{National Research Council, 5071 West Saanich Rd., Victoria, BC, V9E2E7, Canada}


\begin{abstract}
We observed three high-mass star-forming regions  in the W3 high-mass star formation complex with the Submillimeter Array and IRAM 30~m telescope. These regions, i.e. W3 SMS1 (W3 IRS5), SMS2 (W3 IRS4) and SMS3, are in different evolutionary stages and are located within the same large-scale environment, which allows us to study rotation and outflows as well as chemical properties in an evolutionary sense. While we find multiple mm continuum sources toward all regions, these three sub-regions exhibit different dynamical and chemical properties, which indicates that they are in different evolutionary stages. Even within each subregion, massive cores of different ages are found, e.g. in SMS2, sub-sources from the most evolved UCH{\scriptsize II} region to potential starless cores exist within 30 000 AU of each other. Outflows and rotational structures are found in SMS1 and SMS2. Evidence for interactions between the molecular cloud and the H{\scriptsize II} regions is found in the $^{13}$CO channel maps, which may indicate triggered star formation.
\end{abstract}


\keywords{stars: formation -- stars: massive -- ISM: jets and outflows -- ISM: molecules -- ISM: individual (W3)}

\section{Introduction}
The formation and evolution of high-mass stars is still poorly understood. High-mass star formation not only proceeds exclusively in a clustered mode, but it becomes increasingly apparent that several episodes of high- and low-mass star formation can proceed sequentially within the same molecular cloud complex. Well-known examples are NGC 6334I \& I(N) (e.g., \citealt{sandell2000}) or the S255 complex (e.g., \citealt{minier2005, wang2011}). Since the general molecular and stellar environment in these regions is almost the same within their different subregions, the environmental effects are minimized, which allows us to study the high-mass star formation in an evolutionary sense. Although there has been considerable progress in the general understanding of high-mass star formation over the last decade (e.g., \citealt{zinnecker2007,beuther2007}), many questions still remain open. The time evolution of the chemical and physical properties (e.g., rotating/infalling envelope, outflow) is still unclear.

Among the physical properties, massive molecular outflows are among the most studied phenomena. The outflows have been detected in various high-mass star formation stages, from the early stage as IRDCs to relatively evolved high-mass hot cores (e.g., \citealt{beuther2002b, wu2004, zhang2005, beuther2008, fallscheer2009, vasynina2011}). Observations towards the S255 complex show that outflows in the very young and cold region have low velocity and are relatively confined around the driving source, while the outflows associated with the relatively evolved high-mass protostellar objects exhibit large scale collimated high velocity structure \citep{wang2011}. \citet{beuther2005c} proposed an evolutionary sequence for the massive outflows in the high-mass protostellar phase: as the central object continues to accrete, it evolves from a B star to an O star and the outflows become less collimated under the pressure of the increased radiation and stellar wind.

Massive disks are a key component for high-mass stars to overcome radiation pressure and form through disk accretion (e.g., \citealt{krumholz2007, krumholz2009, kuiper2010}). Rotating disks have been detected around intermediate-mass sources ($\sim5-10$~$M_\odot$, e.g., \citealt{zhang1998a, cesaroni2005, cesaroni2007}) but for higher mass objects, only rotating structures known as ``toroids'' \citep{cesaroni2007} have been detected (e.g., \citealt{beltran2004, beuther2005d, fallscheer2009}). These structures are not in Keplerian rotation and have typical radii of thousands of AU. \citet{fallscheer_phd} sees an evolutionary trend that the size of the potential disk signature decreases from IRDCs to hot cores. \citet{wang2011} also detected rotating toroids associated with high-mass protostellar objects and hot cores in the S255 complex, but the evolutionary trend is not clear according to their results.

From IRDCs to high-mass hot cores, massive star formation exhibits diverse chemical properties (e.g., \citealt{sutton1985, schilke1997a, schilke2001, beuther2009, vasynina2011, wang2011}), yet relatively few evolutionary details are known. High resolution observations also show spatial variations among many species (e.g., \citealt{beuther2005a, beuther2009, wang2011}). Several theoretical studies exist on the chemical evolution during massive star formation (e.g., \citealt{charnley1997,nomura2004,viti2004,beuther2009}), however, the high-resolution observational database to test these models is still relatively poor. 

\begin{figure}[htbp]
   \centering
    \includegraphics[angle=0,width=\columnwidth]{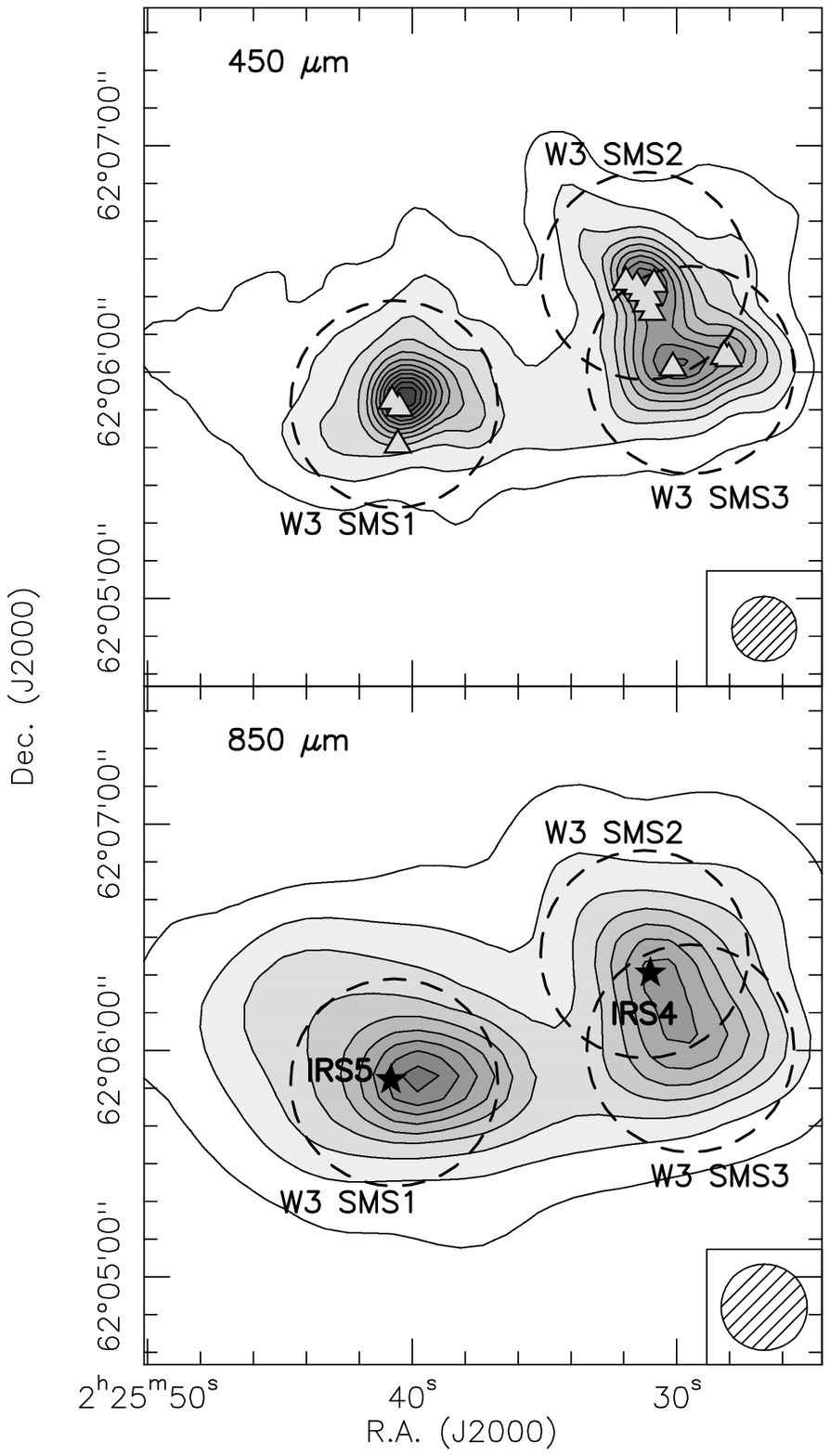}
    \caption{The SCUBA 450~$\mu$m ({\it top}) and 850~$\mu$m ({\it bottom}) continuum images. The contour levels are in steps of 5$\sigma$ starting at 5$\sigma$ in both panels. For the 450 $\mu$m panel, $\sigma=0.1$~Jy~beam$^{-1}$, and for the 850~$\mu$m panel, $\sigma=0.4$~Jy~beam$^{-1}$. The dashed circles mark the primary beam of our SMA observations, the filled triangles in the {\it top panel} mark the SMA continuum sources we resolve, and the stars in the {\it bottom panel} mark the near infrared sources.}
    \label{scuba}
\end{figure}

In this work we present high resolution SMA interferometry observations as well as single-dish CO($2-1$) observations of three subregions (W3 SMS1, W3 SMS2 and W3 SMS3) within the same environment, W3 Main \citep{ladd1993, tieftrunk1998, moore2007, megeath2008}. We study the evolution of the chemical and physical properties (outflows, rotation), ranging from potential starless cores to the most evolved ultracompact H{\scriptsize II} (UCH{\scriptsize II}) regions. W3 Main is part of the W3 molecular cloud and located at a distance of 1.95~$\pm$~0.04~kpc \citep{xu2006}. Single dish sub-mm observations \citep{ladd1993} resolved three continuum sources, W3 SMS1, SMS2 and SMS3 (Fig.~\ref{scuba}).

\paragraph{The easern region W3 SMS1:} The central infrared source W3 IRS5 is associated with the hypercompact H{\scriptsize II} (HCH{\scriptsize II}) cluster W3 M \citep{claussen1994, tieftrunk1997} with a total luminosity of 2$\times10^5$~$L_\odot$ \citep{campbell1995,wynn-williams1972}, and coincides with water and OH masers \citep{claussen1994, sarma2001, sarma2002, imai2000, gaume1987}. The low radio luminosity of these HCH{\scriptsize II} regions suggests they might be stellar wind/jet sources \citep{hoare2007, hoare2007a}, and \citet{wilson2003} found two of the HCH{\scriptsize II} regions show large proper motion and might be jet-knots expelled from the massive stars in IRS5. In the high resolution ($<$ 0.1$''$) radio observations done by \citet{vandertak2005} these sources do not show elongated morphology and the authors argue that there are three HCH{\scriptsize II} regions together with several shock-ionized clumps. \citet{megeath2005} reported seven near-IR sources with HST observations, and three of them have both mid-IR and radio continuum counterparts \citep{vandertak2005}. Using the Plateau de Bure Interferometer, \citet{rodon2008} resolved the main continuum source into five sources. Bipolar outflows and even multiple outflows are also reported \citep{claussen1984, mitchell1992, choi1993, hasegawa1994, imai2000, rodon2008}. It has been proposed that IRS5 is a proto-Trapezium and might emerge as a bound Trapezium similar to that in the Orion nebula \citep{megeath2005, rodon2008}. \citet{megeath1996} identified an embedded cluster of 80$-$240 low-mass stars centered on IRS5, which in turn is surrounded by hundreds of low-mass stars \citep{feiglson2008}. W3 IRS7 (associated with the UCH{\scriptsize II} region W3 F) and IRS3 (associated with the compact H{\scriptsize II} region W3 B) are also found here \citep{wynn-williams1972,harris1976,tieftrunk1997}.

\paragraph{The northwestern region W3 SMS2:} The infrared source W3 IRS4 lies at the center of the continuum source with a luminosity of 6$\times10^4$~$L_\odot$ \citep{campbell1995}. IRS4 is classified as an O8V--B0.5V star and is thought to be the exciting source of the UCH{\scriptsize II} region W3 C \citep{bik2011,tieftrunk1997}. The HCH{\scriptsize II} region W3 Ca lies east of IRS4 which indicates that a massive star is forming. OH masers and evidence for an outflow are also found towards IRS4 \citep{gaume1987, hasegawa1994}.

\paragraph{The southwestern region W3 SMS3:} W3 SMS3, located south of W3 IRS4, is considered a quiescent region in W3 Main \citep{megeath2008}. It exhibits strong mm continuum emission (Fig.~\ref{scuba}) but no other signs of active star formation such as near infrared or mid infrared emission \citep{ruch2007}. \citet{tieftrunk1998} found extended emission towards SMS3 with VLA NH$_3$ observations, which show quite narrow line widths towards this region. \citet{tieftrunk1998} also found four compact NH$_3$ clumps with masses around 30 $M_\odot$ surrounding the main clump.

Furthermore, NIR imaging and spectroscopy studies by \citet{bik2011} revealed an age spread of at least 2--3~Myr in W3 Main and suggested a sequential star formation process. While W3 SMS1 and SMS2 show signs of active star formation and are surrounded by a NIR cluster, SMS3 is still quiescent with no signs of ongoing star formation. Therefore, W3 Main is an ideal region to simultaneously investigate several sites of massive star formation at different evolutionary stages within the same over all environment.

\section{Observations and Data Reduction}

\subsection{Submillimeter Array observations}
The W3 Main complex was observed in three fields with the Submillimeter Array\footnote{The Submillimeter Array is a joint project between the Smithsonian Astrophysical Observatory and the Academia Sinica Institute of Astronomy and Astrophysics and is funded by the Smithsonian Institution and the Academia Sinica.} (SMA) on December 31st 2009 in the compact configuration and on January 30th 2010 in the extended configuration, both with seven antennas. The phase centers of the observations, which are known as W3-SMS1, W3-SMS2, and W3-SMS3, were R.A. 02$^{\rm h}$25$^{\rm m}$40.68$^{s}$ Dec. +62$^\circ$05$'$51.5$''$ (J2000), R.A. 02$^{\rm h}$25$^{\rm m}$31.22$^{\rm s}$ Dec. +62$^{\circ}$06$'$25.5$''$ (J2000), and R.A. 02$^{\rm h}$25$^{\rm m}$29.49$^{\rm s}$ Dec. +62$^\circ$06$'$00.6$''$ (J2000), respectively. The SMA has two spectral sidebands, both 4 GHz wide and separated by 10 GHz. The receivers were tuned to 230.538 GHz in the upper sideband \mbox{($v_{\rm lsr}=-43$~km~s$^{-1}$)} with a spectral resolution of 1.2~km~s$^{-1}$. For the compact configuration, the bandpass calibration was derived from the observations of the quasar 0854+201. Phase and amplitude were calibrated with regularly interleaved observations of the quasar 0136+478 (15.8$^\circ$ from the source). Since the flux measurements of the gain calibrator 0136+478 on December 31st 2009 are available in the SMA calibrator database, we do the flux calibration with this quasar, and the flux scale is estimated to be accurate within 20$\%$. For the extended configuration, the bandpass calibration was derived from the quasar 3C273. Phase and amplitude were calibrated with regularly interleaved observations of the quasar 0102+584 (11$^\circ$ from the source). The flux calibration was derived from observations of Titan, and the flux scale is estimated to be accurate within 20$\%$. We merged the two configuration datasets, applied different robust parameters for the continuum and line data, and obtained synthesized beam sizes between 1.5$''\times1.1''$ (PA --35$^\circ$) and 2.2$''\times1.8''$ (PA --34$^\circ$). The $3\sigma$ rms of the 1.3~mm (225 GHz) continuum image is $\sim$ 10.8~mJy~beam$^{-1}$ and the $3\sigma$ rms of the line data is 0.11 Jy beam$^{-1}$ at 2~km~s$^{-1}$ spectral resolution. The flagging and calibration were done with the IDL superset MIR \citep{scoville1993} which was originally developed for the Owens Valley Radio Observatory and adapted for the SMA\footnote{The MIR cookbook by Chunhua Qi can be found at http://cfa-www.harvard.edu/$\sim$cqi/mircook.html.}. The imaging and data analysis were conducted in MIRIAD \citep{sault1995}. 
 
\subsection{Short spacing from the IRAM 30~m}
To complement the CO($2-1$) observations which lack the short-spacing information and to investigate the large-scale general outflow properties, we observed the three subregions with the HERA array at the IRAM 30~m telescope on February 21st, 2011. The $^{12}$CO$(2-1)$ line at 230.538~GHz and the $^{13}$CO($2-1$) at 220.399~GHz were observed in the on-the-fly mode. We mapped the whole region with a map of 4$'\times4'$ centered at R.A. 02$^{\rm h}$25$^{\rm m}$41.01$^{\rm s}$ Dec. +62$^{\circ}07'02.1''$ (J2000). The sampling interval was 0.8$''$. The region was scanned once along the declination and right ascension directions, respectively, in order to reduce scanning effects. The spectra were calibrated with CLASS, which is part of the GILDAS software package. The $^{12}$CO data have a beam size of 11$''$, and the rms noise level of the corrected $T_{\rm mb}$ scale is around 0.25 K at 1.2~km~s$^{-1}$ spectral resolution. The $^{13}$CO data have the same beam size, but the rms noise is 0.35 K at 1.2~km~s$^{-1}$ spectral resolution.
 
After reducing the 30~m data separately, single-dish $^{12}$CO and $^{13}$CO data were converted to visibilities and then combined with the SMA data using the MIRIAD package task UVMODEL. The synthesized beam of the combined data is 3.0$''\times2.5''$ (PA --34$^\circ$).

\section{Results}
\subsection{Millimeter continuum emission}
\label{continuum}
\begin{figure*}[htbp]
   \centering
    \includegraphics[angle=0,width= 15cm]{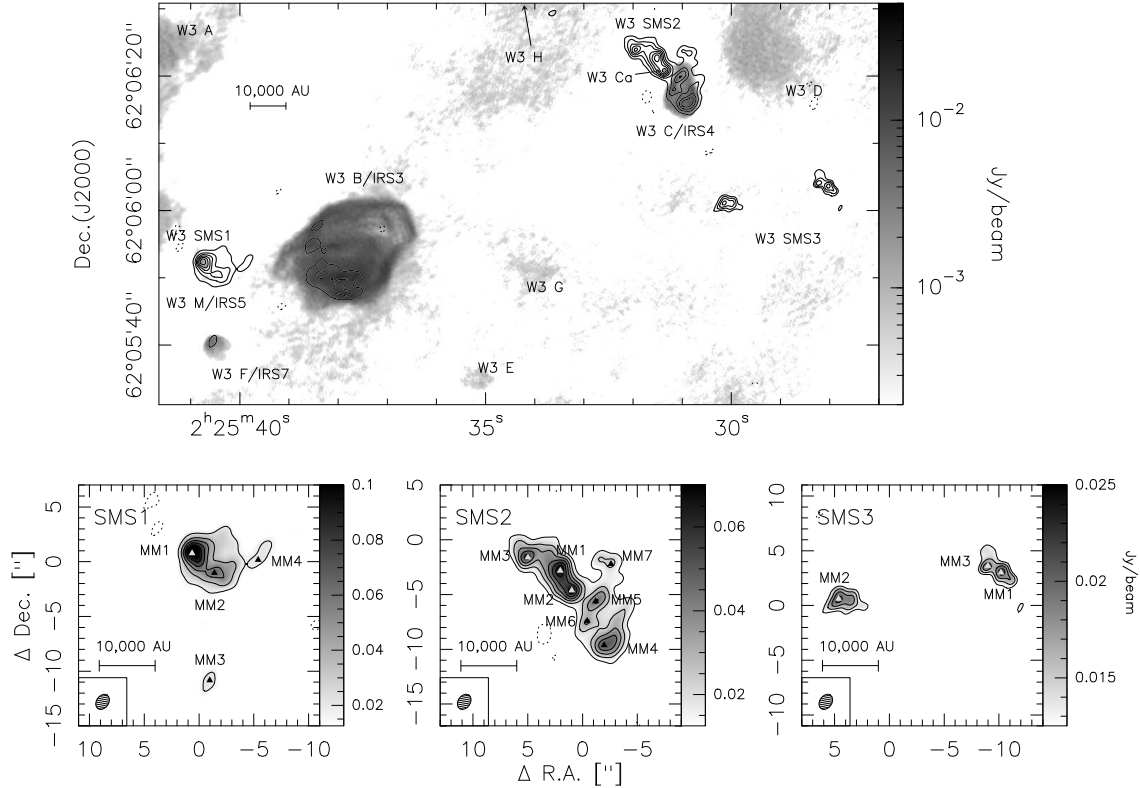}
    \caption{The SMA 1.3~mm continuum maps of W3 SMS1, SMS2 and SMS3 regions. {\it Top panel: } the VLA 22.5~GHz continuum image \citep{tieftrunk1997} overlaid with the SMA 1.3~mm continuum contours. The contour levels start at 5$\sigma$ and are in steps of 5$\sigma$ for SMS1 ($\sigma=$3.6~mJy~beam$^{-1}$), in steps of 4$\sigma$ for SMS2 ($\sigma=$2.5~mJy~beam$^{-1}$) and in steps of 1$\sigma$ in SMS3 region ($\sigma=$2.5~mJy~beam$^{-1}$). {\it Bottom panels:} the SMA 1.3~mm continuum maps of each region. The contour levels are the same as in the {\it top panel}. The dotted contours are negative features due to the missing flux with the same contours levels as the positive ones in each panel. Beams are shown at the bottom left of each panel. Filled triangles mark the millimeter sources detected.}
\label{cont}
\end{figure*}

Figure~\ref{cont} shows the 1.3~mm continuum maps of the three regions. We resolve several continuum sources in all three regions. Assuming optically thin dust emission, we estimate the gas mass and column density of the continuum sources following the equations outlined in \citet{hildebrand1983} and \citet{beuther2005a}. We assume a dust temperature of 40~K, grain size of 0.1~$\mu$m, grain mass density of 3~g~cm$^{-3}$, gas-to-dust ratio of 100, and a grain emissivity index of 2 (corresponding to $\kappa\approx0.3$~cm$^2$~g$^{-1}$ for comparison with \citealt{ossenkopf1994}). Here we assume the same dust temperature for all the continuum sources to simplify the calculation. The fact that the continuum sources may have different temperatures  (see Sect.~\ref{sect_temp}) would also bring uncertainty to the mass results. The SMA continuum measurements listed in Table \ref{conttable} are not corrected for the primary beam, since for SMS1 and SMS2, the primary beam correction would only change the flux of our detected continuum sources by $\leq2\%$, and for SMS3 the flux would be changed by $\leq10\%$. For the SMA continuum sources which are associated with radio emission, the free-free emission is subtracted. We assume optically thin free-free emission (spectral index $\sim-0.1$), and use the VLA 22.5~GHz flux \citep{tieftrunk1997} to estimate the free-free flux at 225 GHz. In the following, we label the mm sources according to a scheme consisting of the name of the large-scale region (SMS1 to SMS3) followed by the mm peaks within them, e.g., SMS1-MM1. The corrected results are shown in Table \ref{conttable}. For sources SMS1-MM3 and SMS2-MM4, the free-free flux we obtain at 225~GHz is even larger than our SMA 1.3~mm measurements, which is likely due to the fact that the VLA observations have better uv coverage and suffer less from missing flux than the SMA observations. 

With the same set-up, we also calculate the gas mass and the column density of the SCUBA 850~$\mu$m continuum peaks shown in Table \ref{scubatable} (The SCUBA data were downloaded from the archive, \citealt{difrancesco2008}) and compare the results with the ones in Table \ref{conttable}. The masses we obtain from the SMA observations for the W3 SMS1, W3 SMS2 and W3 SMS3 are only $9\%$, $12\%$ and $3\%$ of the SCUBA 850~$\mu$m measurements, respectively. For the W3 SMS1 region, we resolve part of the H{\scriptsize II} region W3 B from the SMA observations (Fig.~\ref{cont}). However, it is already near the edge of our SMA field of view, and the emission we recover should be dominated by free-free emission. Therefore, we do not include the emission at W3 B from the SMA observations for the comparison. Since our interferometer observations are not sensitive to spatial scales $>27''$ (corresponding to the shortest baseline $=7.5$~k$\lambda$), we filter out the smoothly distributed large-scale halo and observe only the compact cores. Furthermore, the observational fact that we filter out more flux for the youngest region W3 SMS3 indicates that at the earliest evolutionary stages the gas is more smoothly distributed than during later stages where the collapse produces more centrally condensed structures. Nevertheless, all of the column densities we derive are above the estimated threshold for high-mass star formation of 1~g~cm$^{-2}$ \citep{krumholz2008}, which corresponds to a column density of $N_{\rm H_2}\sim3\times10^{23}$~cm$^{-2}$.

\begin{deluxetable*}{lccccccccc} 
\tablewidth{0cm}
\tablecaption{Millimeter continuum peak properties.}
\setlength{\tabcolsep}{0.05cm}
\tabletypesize{\footnotesize}
\centering
\tablehead{
\colhead{Source}&
\colhead{R.A.}&
\colhead{Dec.}&
\colhead{$I_\nu$}&
\colhead{$S_\nu$}&
\colhead{Mass}&
\colhead{$N_{\rm H_2}$}&
\colhead{free-free $I_\nu$}&
\colhead{}&
\colhead{free-free $S_\nu$}\\
\cline{8-8}
\cline{10-10}
\colhead{}&
\colhead{(J2000)}&
\colhead{(J2000)}&
\colhead{(mJy/beam)}&
\colhead{(mJy)}&
\colhead{($M_\odot$)}&
\colhead{($10^{24}$ cm$^{-2}$)}&
\colhead{measured $I_\nu$}&
\colhead{}&
\colhead{measured $S_\nu$}
}
\startdata
SMS1-MM1&02:25:40.77&+62:05:52.3&         123&405&          40&6.5&0.03&&  0.02\\
SMS1-MM2&02:25:40.48&+62:05:50.5&          63&229&          23&3.4&n/a&& n/a\\
SMS1-MM3&02:25:40.54&+62:05:40.7&          22&25&           -&0.3&0.73&&  2.04\\
SMS1-MM4&02:25:39.91&+62:05:51.7&          22&35&           4&1.2&n/a&& n/a\\
\hline
\noalign{\smallskip}
SMS2-MM1&02:25:31.51&+62:06:22.7&          68&208&          21&3.6&n/a&&  n/a\\
SMS2-MM2&02:25:31.36&+62:06:20.9&          67&121&          12&3.5&0.02&&  0.01\\
SMS2-MM3&02:25:31.93&+62:06:23.9&          55&113&          11&2.9&n/a&& n/a\\
SMS2-MM4&02:25:30.94&+62:06:15.9&          50&220&           -&1.3&0.52&& 1.01\\
SMS2-MM5&02:25:31.05&+62:06:19.9&          43&75&           2&1.3&0.42&&0.74\\
SMS2-MM6&02:25:31.16&+62:06:18.1&          34&50&           0.1&0.6&0.69&&  0.97\\
SMS2-MM7&02:25:30.85&+62:06:23.3&          26&67&           7&1.4&n/a&& n/a\\
\hline
\noalign{\smallskip}
SMS3-MM1&02:25:28.04&+62:06:03.6&          21&40&           4&1.2&n/a&&  n/a\\
SMS3-MM2&02:25:30.15&+62:06:01.2&          21&66&           7&1.2&n/a&&  n/a\\
SMS3-MM3&02:25:28.21&+62:06:04.2&          19&33&           3&1.1&n/a&&  n/a
\enddata
\tablecomments{``n/a'' indicates no free-free emission is detected towards this source. For sources SMS1-MM3 and SMS2-MM4, the free-free flux we obtain at 225 GHz is even larger than our SMA 1.3 mm measurements, so we do not have the mass estimation.}
\label{conttable}
\end{deluxetable*}

\begin{deluxetable}{lcccc}
\tablewidth{0cm}
\tablecaption{SCUBA 850~$\mu${\rm m} continuum data.}
\setlength{\tabcolsep}{0.1cm}
\tabletypesize{\footnotesize}
\centering
\tablehead{
\colhead{Source}&
\colhead{$I_\nu$}&
\colhead{$S_\nu$}&
\colhead{Mass}&
\colhead{$N_{\rm H_2}$}\\
\colhead{}&
\colhead{(Jy/beam)}&
\colhead{(Jy)}&
\colhead{($M_\odot$)}&
\colhead{(10$^{23}$ cm$^{-2}$)}
}
\startdata
   W3 SMS1 &19 & 39 & 715 & 5.9\\
   W3 SMS2 &15 & 24 & 435 & 4.7\\
   W3 SMS3 &14 & 24 & 445 & 4.4
\enddata
\label{scubatable}
\end{deluxetable}

\begin{deluxetable}{lccc}
\tablewidth{0cm}
\tablecaption{ Millimetric, radio, NIR and MIR counterparts.}
\setlength{\tabcolsep}{0.1cm}
\tabletypesize{\footnotesize}
\centering
\tablehead{
\colhead{MM}&
\colhead{radio\tablenotemark{a}}&
\colhead{NIR}&
\colhead{MIR}
}
\startdata
SMS1-MM1&	W3 M (a+b+c+d1+d2+f)&	IRS5\tablenotemark{b}&	1+2\tablenotemark{c}\\
SMS1-MM3&	W3 F			  & IRS7\tablenotemark{b}&   ... \\
SMS2-MM2&	W3 Ca			  & IRS4-b\tablenotemark{d}&	IRS4-b\tablenotemark{d}\\
SMS2-MM4&	W3 C			  & ...   & ... \\
SMS2-MM5&	W3 C              & IRS4-a\tablenotemark{d}& ...\\
SMS2-MM6&   W3 C              & ...   &...
\enddata
\tablenotetext{a}{\citet{claussen1994,tieftrunk1997}}
\tablenotetext{b}{\citet{wynn-williams1972}}
\tablenotetext{c}{\citet{vandertak2005}}
\tablenotetext{d}{Figs. \ref{sms2_ks} and \ref{sms2_ch4}}
\label{counterparts}
\end{deluxetable}

\paragraph{W3 SMS1}
The SMA continuum image of W3 SMS1 is shown in the bottom left panel of Fig. \ref{cont}. The SMA image reveals four continuum sources, SMS1-MM1, SMS1-MM2, SMS1-MM3 and SMS1-MM4 in this region. As shown in Table \ref{counterparts}, the strongest source SMS1-MM1 coincides with the near-infrared source W3 IRS5 and the HCH{\scriptsize II} region cluster W3 M \citep{claussen1994,tieftrunk1997}. It is also associated with the mid-infrared sources MIR1 and MIR2 detected by \citet{vandertak2005}. H$_2$O masers are also detected at the SMS1-MM1 position \citep{imai2000, sarma2001, sarma2002}. A previous higher resolution study shows that SMS1-MM1 is resolved into 4 individual 1.4~mm continuum sources \citep{rodon2008}. SMS1-MM2 and MM4, which have not been detected before, are not associated with any infrared source. SMS1-MM4 is close to an NH$_3$ core detected by \citet{tieftrunk1998}. SMS1-MM3, south of SMS1-MM1, is associated with the infrared source W3 IRS7 and the UCH{\scriptsize II} region W3 F, Table \ref{conttable} shows that the continuum emission at this region is dominated by free-free emission from the UCH{\scriptsize II} region and it is relatively evolved \citep{tieftrunk1997}.

\begin{figure}[htbp]
   \centering
    \includegraphics[angle=0,width=\columnwidth]{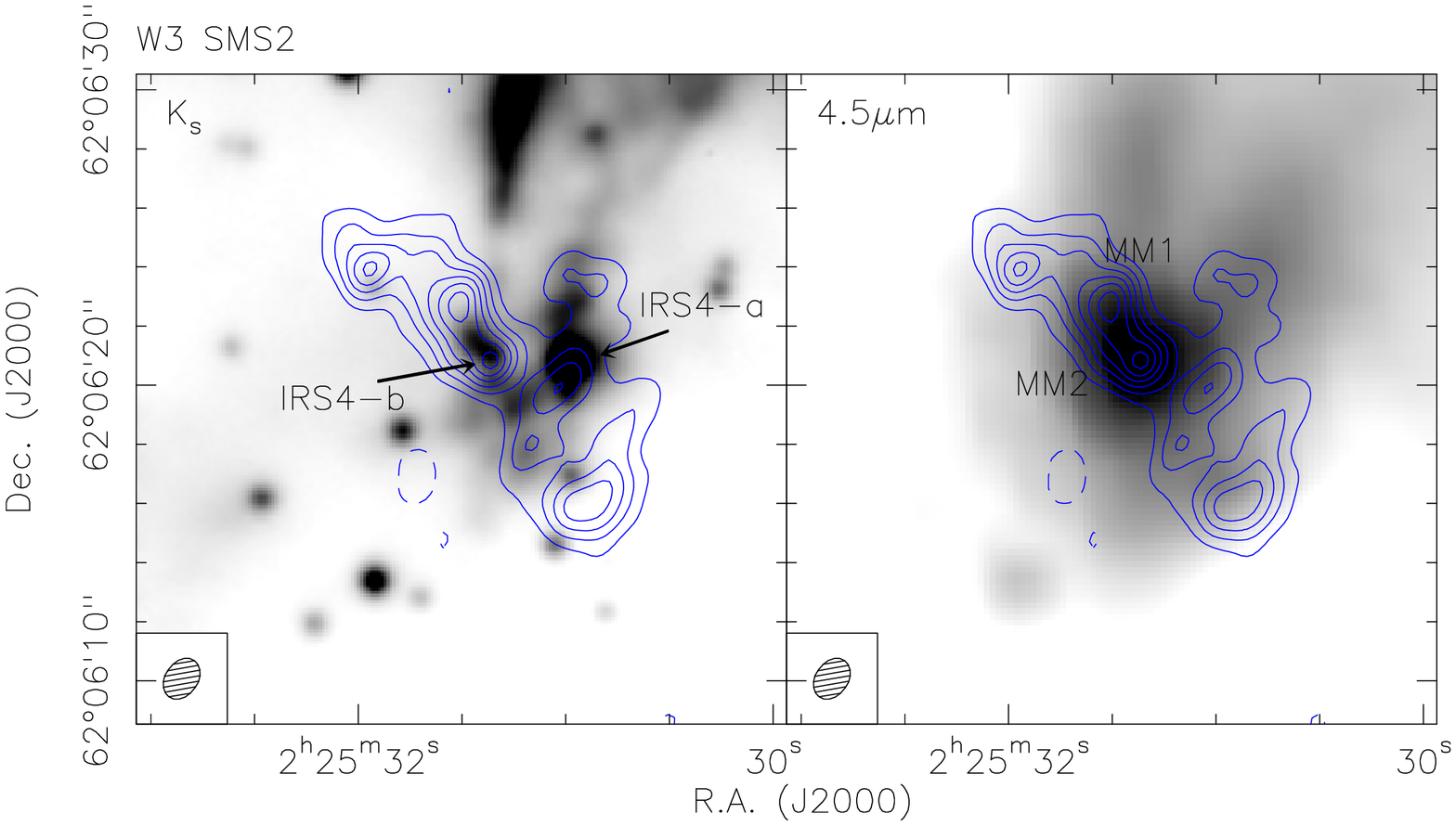}
  \caption{{\it Left panel:} The LBT-LUCI $K_s$-band image overlaid with the SMA 1.3~mm continuum map of the W3 SMS2 region. The $K_s$-band image is adapted from \citet{bik2011}. {\it Right panel:} The IRAC 4.5 $\mu$m short exposure time image \citep{ruch2007} overlaid with the SMA 1.3~mm continuum image. The IRAC post-bcd data were processed with pipeline version S18.7.0 and downloaded from the {\it Spitzer} archive. The contour levels are in steps of 4$\sigma$ starting at 5$\sigma$ ($\sigma=2.5$~mJy~beam$^{-1}$). The dashed contours are the negative features due to the missing flux with the same contours levels as the positive ones in each panel. The beam of the SMA map is shown at the bottom left corner.}
\label{sms2_ks}
\end{figure}

\paragraph{W3 SMS2}
The bottom middle panel of Fig.~\ref{cont} presents the SMA continuum image of W3 SMS2. We detect 7 continuum sources in this region. As listed in Table \ref{counterparts}, SMS2-MM2 is associated with the HCH{\scriptsize II} region W3 Ca \citep{tieftrunk1997}, and Fig.~\ref{sms2_ks} shows that SMS2-MM2 also coincides with another elongated infrared source IRS4-b which could be tracing the jet driven by SMS2-MM2 (see Fig.~\ref{sms2_out} in Sect. \ref{outflow}). The elongated infrared source IRS4-b is very luminous in {\it Spitzer}/IRAC bands and dominates the mid-infrared emission here. The NIR source IRS4 marked here as IRS4-a fades out (Fig.~\ref{sms2_ks}, \citealt{ruch2007}). The top panel of Fig.~\ref{cont} shows that SMS2-MM4, SMS2-MM5, SMS2-MM6 are associated with the ultracompact drop-shaped H{\scriptsize II} region W3 C, and Table \ref{conttable} shows that the continuum emission of these three sources is dominated by the free-free emission from the UCH{\scriptsize II} region W3 C. This indicates that SMS2-MM4, SMS2-MM5 and SMS2-MM6 are just part of the UCH{\scriptsize II} region W3 C. The other three continuum sources, SMS2-MM1, MM3 and MM7, are not associated with any infrared or radio continuum source. They are likely to be at a very young evolutionary stage.
     
\paragraph{W3 SMS3}
The SMA continuum image of W3 SMS3 is shown in the bottom right panel in Fig.~\ref{cont}. The SMA observations resolve three continuum peaks in this region. These continuum sources are not associated with any near infrared or radio source. The {\it Spitzer}/MIPS 24 $\mu$m map also shows no emission toward this region \citep{ruch2007}. Meanwhile, the gas mass is only 3$\%$ of the SCUBA 850~$\mu$m measurements, which indicates that the gas in this region is smoothly distributed indicative of an early evolutionary stage. All these features indicate that W3 SMS3 is an extremely young region.

\subsection{Spectral line emission}
\label{spectra}
\begin{figure}[htbp]
   \centering
    \includegraphics[angle=0,width= \columnwidth]{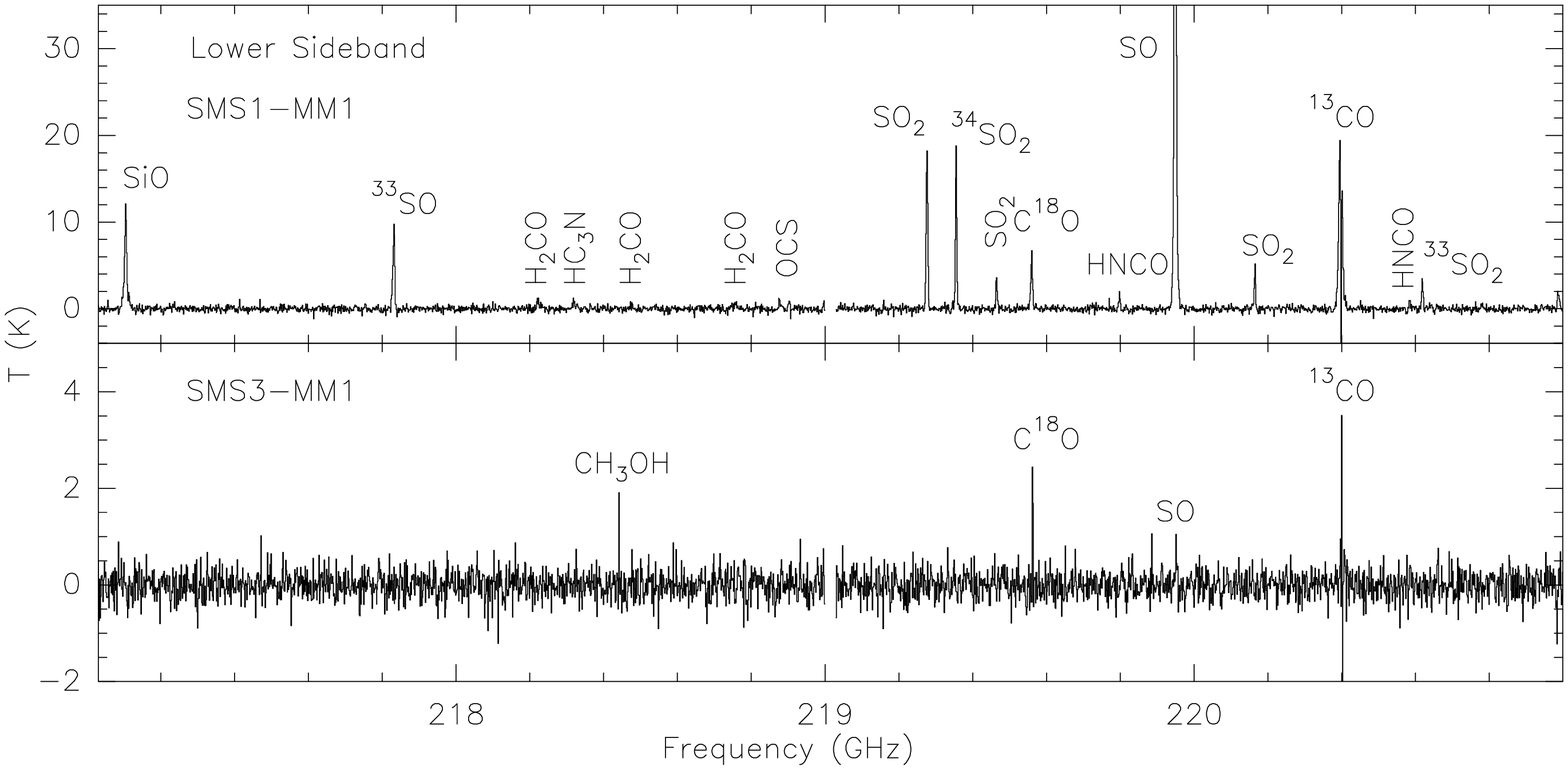}\\
    \includegraphics[angle=0,width= \columnwidth]{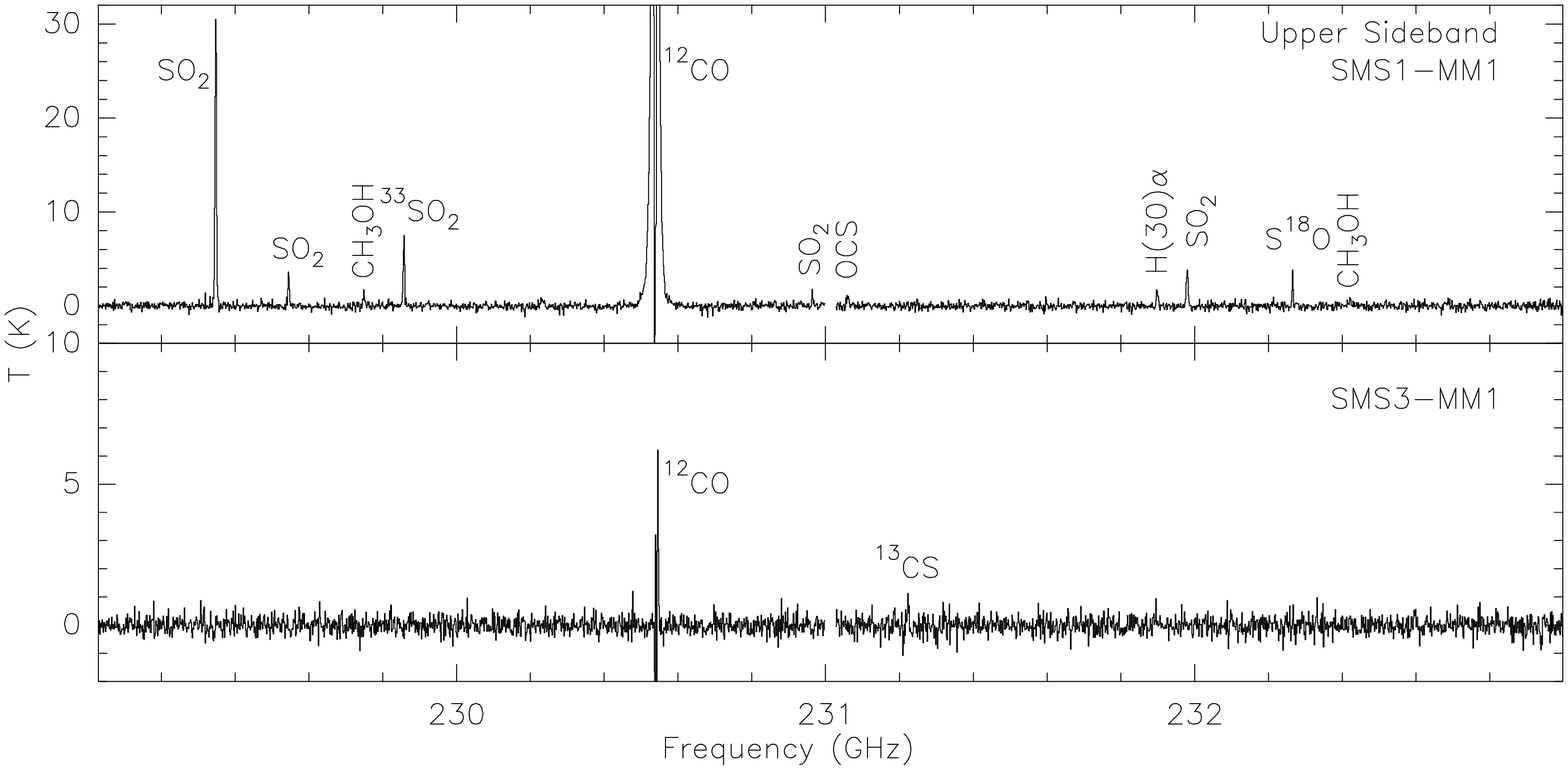}
  \caption{{\it Upper} and {\it lower} sideband spectra extracted towards SMS1-MM1 and SMS3-MM1 with a resolution of 2~km~s$^{-1}$ per channel. The SO line and the $^{12}$CO line in SMS1-MM1 are not fully plotted, which go to 84 K and 94 K, respectively. The small gap in each spectrum is an effect of the data reduction process, since each sideband is divided into two parts to be processed separately.}
\label{spec_fig}
\end{figure}

We detect 41 lines from 13 species towards all three regions. In addition to the three CO isotopologues $^{12}$CO, $^{13}$CO and C$^{18}$O, we also detect dense/hot gas tracers CH$_3$CN, CH$_3$OH, HNCO, HC$_3$N, OCS, shock tracer SiO, temperature determinator H$_2$CO, ionized gas tracer hydrogen recombination line H30$\alpha$, and many SO and SO$_2$ isotopologues. The detailed chemical properties of each region will be explained in the following sections. We extract spectra of several molecular transitions at the position of each continuum peak within one beam size ($2.1''\times1.7''$ for the lower sideband and $2.0''\times1.6''$ for the upper sideband), and the detected lines are listed in Table \ref{spectable}. Figure~\ref{spec_fig} shows the spectra towards SMS1-MM1 and SMS3-MM1 as examples.

\paragraph{W3 SMS1}
In the W3 SMS1 region, we detect 36 lines from 12 species. Besides the three CO isotopologues, we also detect many sulfur-bearing species ($^{13}$CS, OCS, SO, SO$_2$ and their isotopologues), some dense gas molecules which are usually used to trace high-mass hot cores, such as CH$_3$OH, CH$_3$CN \citep{nomura2004, beuther2009, sutton1985}, and the well known kinetic temperature determinator H$_2$CO \citep{mangum1993, jansen1994, muehle2007, watanabe2008}. All the lines we detect have lower energy levels $E_{\rm lower}/k$ between 5.3 to 881 K (Table \ref{spectable}).

\begin{figure*}[htbp]
   \centering
    \includegraphics[angle=0,width= 14cm]{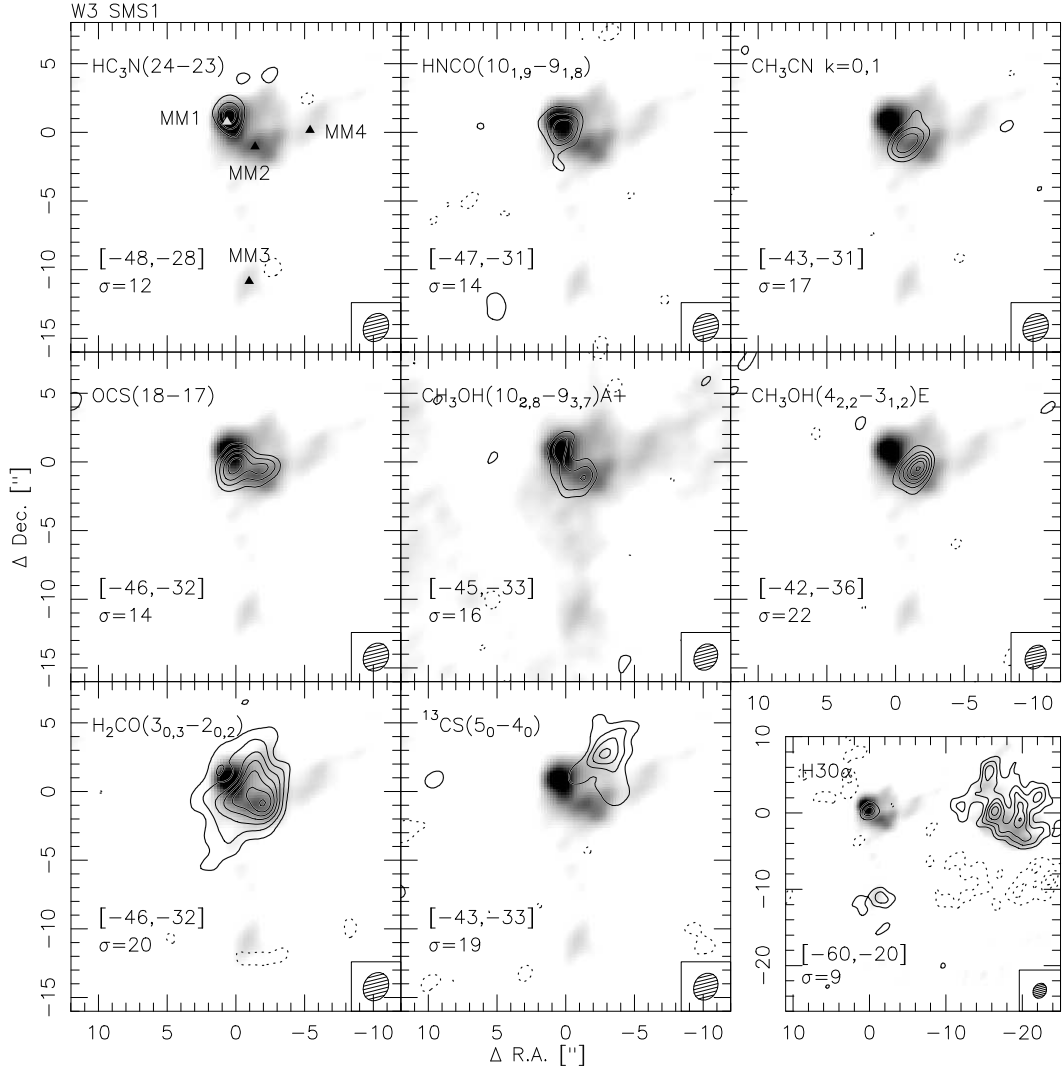}
    \caption{W3 SMS1 molecular line integrated intensity maps overlaid on the SMA 1.3~mm continuum emission in the background. Contour levels start at 3$\sigma$ and continue in steps of 2$\sigma$. The $\sigma$ value for each transition is shown in the respective map in mJy~beam$^{-1}$. The dotted contours are the negative features due to the missing flux with the same contour levels as the positive ones in each panel. The integrated velocity ranges are shown in the bottom left part of each panel in km~s$^{-1}$. The synthesized beams of the molecular line integrated intensity images are shown in the bottom right of each panel. The (0, 0) point in each panel is R.A. 02$^{\rm h}$25$^{\rm m}$40.68$^{\rm s}$ Dec. $+62^{\circ}$05$'$51.5$''$ (J2000).}
\label{sms1_others}
\end{figure*}

\begin{figure*}[htbp]
   \centering
    \includegraphics[angle=0,width= 14cm]{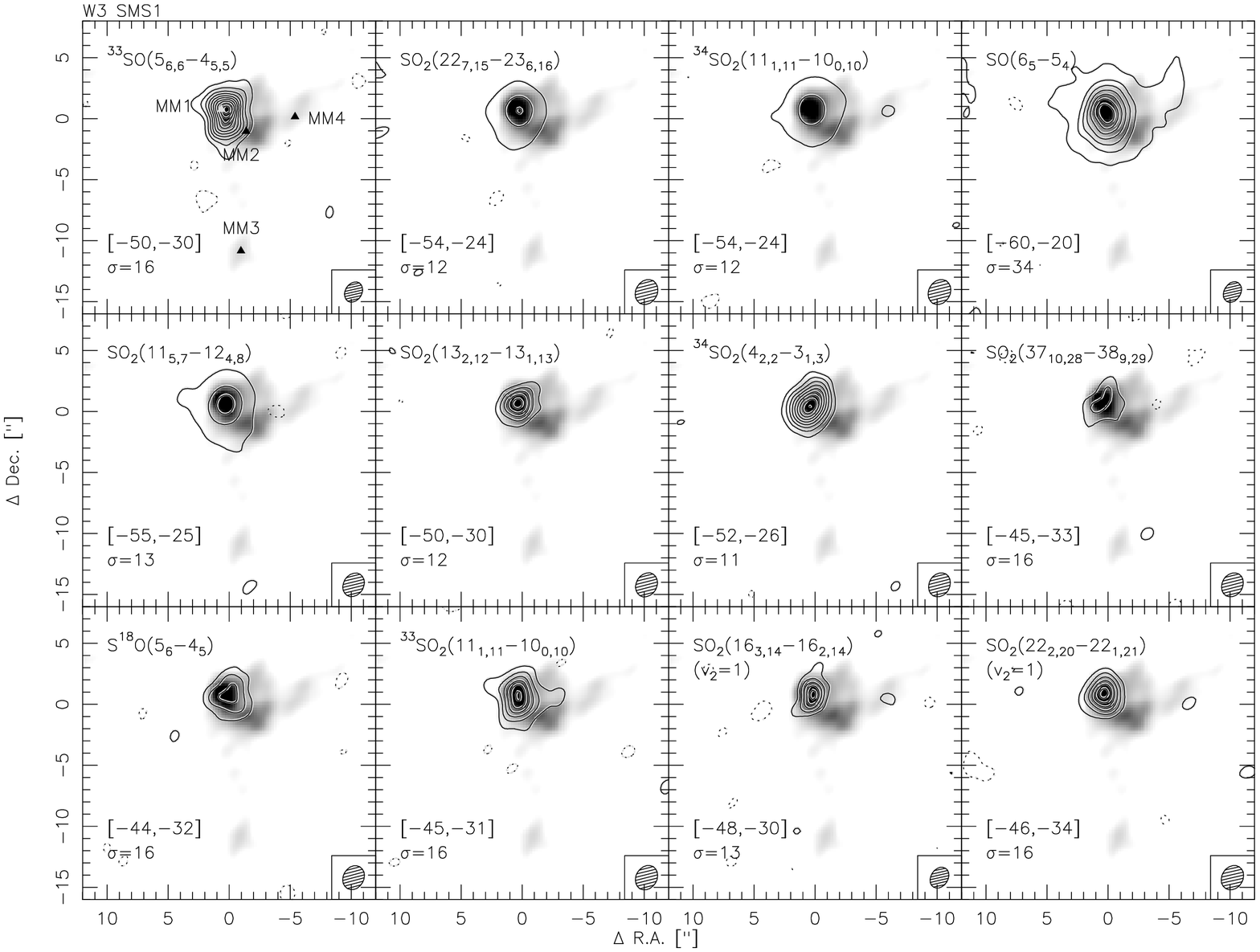}
    \caption{W3 SMS1 sulfur-oxides line integrated intensity map overlaid on the SMA 1.3~mm continuum emission in the background. Contour levels start at 3$\sigma$ and continue in steps of 3$\sigma$, except for the SO($6_5-5_4$) panel, in which the contour step is 10$\sigma$. The $\sigma$ value for each transition is shown in the respective map in mJy~beam$^{-1}$. The dotted contours are the negative features due to the missing flux with the same contour levels as the positive ones in each panel. The integrated velocity ranges are shown in the bottom left part of each panel in~km~s$^{-1}$. The synthesized beams of the molecular line integrated intensity images are shown in the bottom right of each panel. The (0, 0) point in each panel is R.A.~02$^{\rm h}$25$^{\rm m}$40.68$^{\rm s}$ Dec.~$+62^{\circ}$05$'$51.5$''$ (J2000).}
\label{sms1_so}
\end{figure*}

Figures \ref{sms1_others} and \ref{sms1_so} present the integrated line maps of all species (except SiO, $^{12}$CO and $^{13}$CO which will be discussed in detail in Sect. \ref{outflow}). For CH$_3$CN($12_k-11_k$), we only present the line map of $k=0,1$, since the other line maps are similar to the one shown. For CH$_3$OH, we only present the line maps of two transitions, and for H$_2$CO we only show the map for one transition. The CH$_3$OH($10_{2,8}-9_{3,7}$)A+ line map shows an elongated structure closely aligned with the direction of outflow-a (see Fig. \ref{sms1_out}, Sect. \ref{outflow}). All the other CH$_3$OH and CH$_3$CN lines do not peak exactly on any continuum source but somewhere between SMS1-MM1 and SMS1-MM2 (Fig.~\ref{sms1_others}). We derive the peak positions of the CH$_3$CN($k=0,1$), CH$_3$OH(4$_{2,2}-3_{1,2}$)E and H30$\alpha$ integrated emission through Gaussian fitting, and list the results in Table \ref{line_pos}. All three H$_2$CO lines show an elongated structure which extends in the northwest-southeast direction centered between the SMA continuum peaks SMS1-MM1 and SMS1-MM2. $^{13}$CS line emission is offset from the SMA continuum peaks (Fig.~\ref{sms1_others}), similar offset C$^{34}$S emission has also been found by \citet{beuther2009}. The hydrogen recombination line H30$\alpha$ is also detected towards SMS1-MM1, SMS1-MM3 and the H{\scriptsize II} region W3 B. Table \ref{line_pos} shows that the H30$\alpha$ peak at SMS1-MM1 is between the compact radio sources a and b \citep{claussen1994, tieftrunk1997}, and the emission might be the joint contribution of sources a, b, c, d1/d2 and f \citep{claussen1994,tieftrunk1997}. We detect many sulfur-bearing lines in this region. Figure~\ref{sms1_so} shows that all lines of SO and SO$_2$ isotopologues peak on the continuum source SMS1-MM1. Furthermore, the SO($6_5-6_4$) line emission shows an elongated structure along the red-shifted outflow (outflow-b in Fig.~\ref{sms1_out}, see Sect. \ref{outflow}) and could be affected by the outflow. 

Most of the lines we detected in the SMS1 region are found towards SMS1-MM1, except the $k$-ladder lines of CH$_3$CN and the $^{13}$CS lines. Far fewer lines are detected towards SMS1-MM2. Although the hot core tracers CH$_3$CN and CH$_3$OH are detected toward SMS1-MM2, the line emission peaks do not coincide with the continuum peak of SMS1-MM2. In addition, CH$_3$OH emission is also associated with the red-shifted outflow of SMS1-MM1(outflow-a in Fig.~\ref{sms1_out}, see Sect. \ref{outflow}), these lines are more likely due to the shock heating excited by the outflow. As a typical dense gas molecule, we cannot explain why no CH$_3$CN emission is detected towards SMS1-MM1. Many lines we label as detected towards SMS1-MM2 in Table \ref{spectable} do not show an emission peak at SMS1-MM2 (Figs. \ref{sms1_others} and \ref{sms1_so}). The reason could be that some of the lines associated with SMS1-MM1 are actually extended emission and not confined to SMS1-MM1. Since we extract the spectra with one beam size, the spectra we extract towards SMS1-MM2 could be affected by the extended emission around SMS1-MM1, which means the actual number of lines which originate from SMS1-MM2 is even smaller. For SMS1-MM4, only the three CO lines, the SO($6_5-5_4$) line and the SiO line are detected (Fig.~\ref{sms1_out}, see Sect. \ref{outflow}). SiO is a well-known shock tracer (e.g., \citealt{schilke1997}), and the SiO emission detected from SMS1-MM4 is clearly due to the shock excited by the outflow. Therefore, SMS1-MM2 and SMS1-MM4 appear to be chemically younger than SMS1-MM1. No lines are detected towards SMS1-MM3 except CO and H30$\alpha$ lines which indicates that this source is relatively evolved and has no high density gas in its circumstellar environment.

\begin{deluxetable}{lcc}
\tablewidth{0cm}
\tablecaption{CH$_3$CN, CH$_3$OH and H30$\alpha$ peak positions.}
\setlength{\tabcolsep}{0.02cm}
\tabletypesize{\footnotesize}
\centering
\tablehead{
\colhead{lines}&
\colhead{R.A. (J2000)}&
\colhead{Dec. (J2000)}
}
\startdata
CH$_3$CN($k=1,2$)     & 2:25:40.54$\pm$0.02  & +62:05:50.9$\pm$0.3\\
CH$_3$OH(4$_{2,2}-3_{1,2}$)E	  & 2:25:40.47$\pm$0.01  & +62:05:51.0$\pm$0.2\\
H30$\alpha$\tablenotemark{a}   & 2:25:40.68$\pm$0.03  & +62:05:51.7$\pm$0.5
\enddata
\tablenotetext{a}{We only fitted the peak at SMS1-MM1.}
\label{line_pos}
\end{deluxetable}

\begin{figure*}[htbp]
   \centering
    \includegraphics[angle=0,width= 14cm]{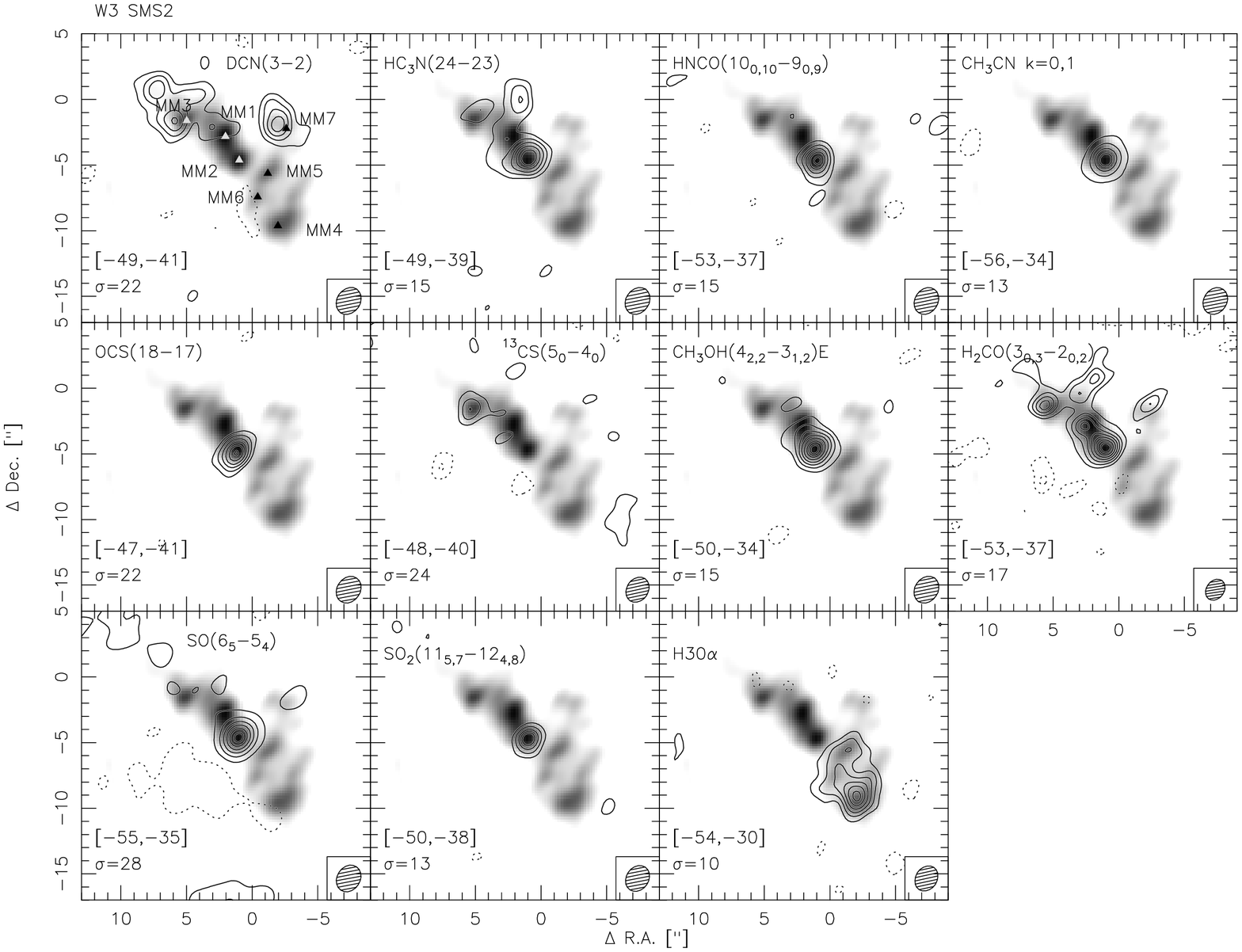}
    \caption{W3 SMS2 molecular line integrated intensity images overlaid on the SMA 1.3~mm continuum emission in the background. Contour levels start at 3$\sigma$ and continue in steps of 2$\sigma$ except for the CH$_3$CN and SO($6_5-5_4$) panels, in which the contour step is 5$\sigma$/level and 10$\sigma$/level, respectively . The $\sigma$ value for each transition is shown in the respective map in mJy~beam$^{-1}$. The dotted contours are the negative features due to the missing flux with the same contour levels as the positive ones in each panel. The integrated velocity ranges are shown in the bottom left part of each panel in~km~s$^{-1}$. The synthesized beams of the molecular line integrated intensity images are shown in the bottom right of each panel. The (0, 0) point in each panel is R.A.~02$^{\rm h}$25$^{\rm m}$31.22$^{\rm s}$ Dec.~$+62^{\circ}$06$'$25.52$''$ (J2000).}
\label{sms2_others}
\end{figure*}

\paragraph{W3 SMS2}
Here we detect 27 molecular lines from 11 species ($^{12}$CO, DCN, H$_2$CO, HC$_3$N, CH$_3$OH, OCS, HNCO, SO, SO$_2$, $^{13}$CS, CH$_3$CN) and 2 additional CO isotopologues ($^{13}$CO and C$^{18}$O) with lower energy levels $E_{\rm lower}/k$ between 5.3 K and 315 K (Table \ref{spectable}). The hydrogen recombination line H30$\alpha$ is also detected towards SMS2-MM4, MM5 and MM6. Figure \ref{sms2_others} shows the integrated line maps of all species (except the CO isotopologues which will be discussed in Sect. \ref{outflow}).

Figure \ref{sms2_others} shows that the hot and dense gas tracers, such as CH$_3$CN and CH$_3$OH, all peak on the SMA continuum source SMS2-MM2, which indicates the massive hot core nature of this continuum source. The H$_2$CO line also peaks at SMS2-MM2 and exhibits extended emission towards SMS2-MM1, MM3 and MM7. Other detected lines at SMS2 all peak at SMS2-MM2 except for DCN and $^{13}$CS. The DCN line shows extended emission towards SMS2-MM1, MM3 and MM7, and the $^{13}$CS line peaks at SMS2-MM3 and extends towards SMS2-MM1. All these features show that the continuum sources SMS2-MM1, MM3 and MM7 are at a younger evolutionary stage than SMS2-MM2. The detection of the H30$\alpha$ line emission towards SMS2-MM4, MM5 and MM6 shows again that these three continuum sources are part of the UCH{\scriptsize II} region W3~C.

\begin{figure}[htbp]
   \centering
    \includegraphics[angle=0,width= \columnwidth]{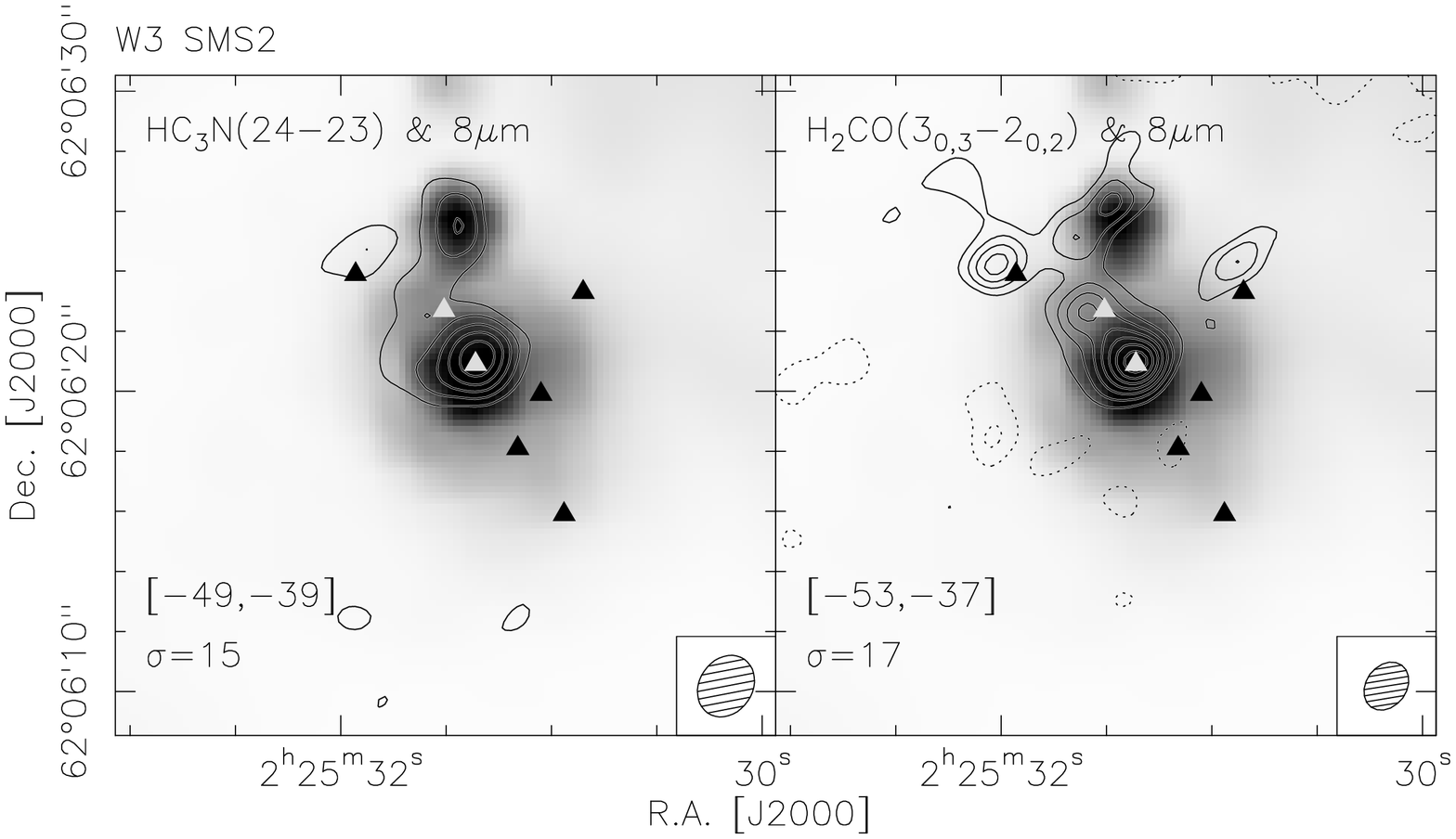}
	\caption{The IRAC 8 $\mu$m short exposure time image \citep{ruch2007} overlaid with the SMA HC$_3$N (left) and H$_2$CO (right) integrated intensity images. The filled triangles mark the SMA millimeter sources we detect. The IRAC post-bcd data were processed with pipeline version S18.7.0 and downloaded from the {\it Spitzer} archive. Contour levels start at 3$\sigma$ and continue in steps of 2$\sigma$. The $\sigma$ value for each transition is shown in the respective map in mJy~beam$^{-1}$. The dotted contours are the negative features due to the missing flux with the same contour levels as the positive ones in each panel. The synthesized beams of the molecular line integrated intensity images are shown in the bottom right of each panel.}
	\label{sms2_ch4}
\end{figure}

The HC$_3$N and H$_2$CO integrated intensity maps show an extra peak to the north of the SMA continuum source SMS2-MM1 (Fig.~\ref{sms2_others}), and one IRAC 8 $\mu$m point source is found there (Fig.~\ref{sms2_ch4}). We name this 8 $\mu$m point source as IRS4-c, which is not visible at shorter wavelengths and does not have a mm continuum counterpart. These features show the peculiar properties of IRS4-c. 

\begin{figure*}[htbp]
   \centering
    \includegraphics[angle=0,width= 14cm]{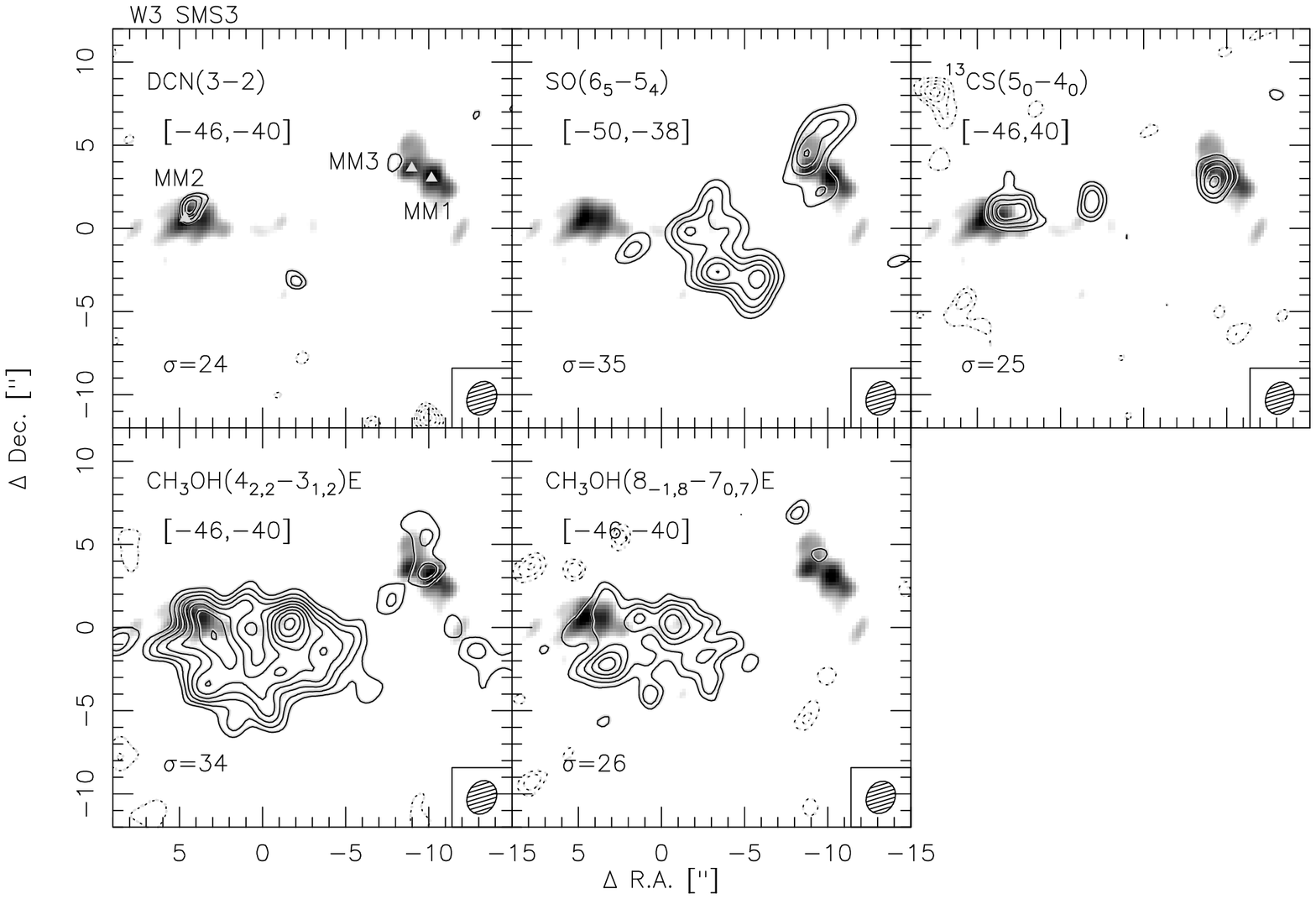}
    \caption{W3 SMS3 molecular line integrated intensity images overlaid on the SMA 1.3~mm continuum emission. Contour levels start at 3$\sigma$ and continue in steps of 1$\sigma$ except for the DCN(3$-$2) panel, in which the contour step is 0.5$\sigma$/level. The $\sigma$ value for each transition is shown in the respective map in mJy~beam$^{-1}$. The dotted contours are the negative features due to the missing flux with the same contour levels as the positive ones in each panel. The integrated velocity ranges are shown in the bottom left part of each panel in~km~s$^{-1}$. The synthesized beams of the molecular line integrated intensity images are shown in the bottom right of each panel. The (0, 0) point in each panel is R.A.~02$^{\rm h}$25$^{\rm m}$29.49$^{\rm s}$ Dec.~$+62^{\circ}06'00.6''$ (J2000).}
\label{sms3_others}
\end{figure*}

\paragraph{W3 SMS3}
Far fewer lines were observed in the W3 SMS3 region. We detect 7 lines from 5 species ($^{12}$CO, DCN, CH$_3$OH, SO and $^{13}$CS) and 2 additional CO isotopologues ($^{13}$CO and C$^{18}$O) with lower energy levels $E_{\rm lower}$ between 5.3 K and 35 K (Table \ref{spectable}). Compared to the other two regions, the excitation temperatures of the lines we detect in this region are much lower, which indicates that W3 SMS3 has a much lower temperature than the other two. Figure \ref{sms3_others} shows the integrated line images of all species (except the three CO isotopologues which will be discussed in Sect. \ref{outflow}). The DCN line emission peaks at SMS3-MM2 and the $^{13}$CS line peaks at SMS3-MM1 and MM2. The CH$_3$OH and SO also peak towards the region between SMS3-MM1 and SMS3-MM2 and exhibit extended emission in the direction of one of the continuum sources. Because the continuum emission is usually more compact than most molecular emission from SO or CH$_3$OH, missing flux is unlikely to explain the discrepancy. The extended gas emission in regions without detected continuum rather indicates a peculiar chemistry. The small number of lines detected in SMS3 indicates that SMS3 is chemically younger than SMS1 and SMS2.

\subsection{The molecular outflows}
\label{outflow}
\begin{figure*}[htbp]
   \centering
    \includegraphics[angle=0,width= 15cm]{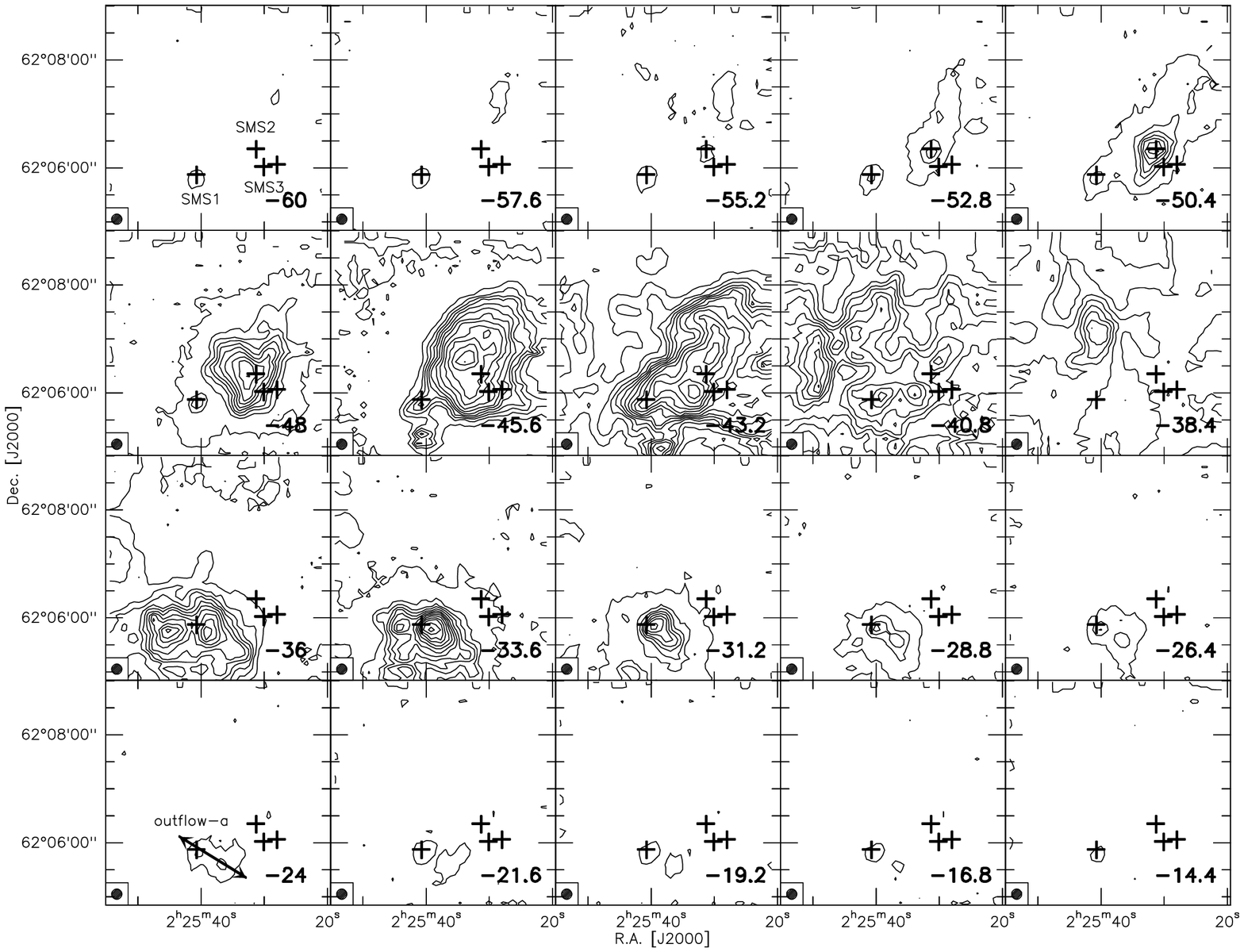}
    \caption{The single dish $^{12}$CO channel map with a spectral resolution of 2.4~km~s$^{-1}$. The contour levels start at 5$\sigma$ and continue in steps of 30$\sigma$ (1$\sigma=0.2$ K). The crosses mark the positions of the SMA continuum sources SMS1-MM1, SMS2-MM2, SMS3-MM1 and SMS3-MM2. The arrow in the bottom left panel marks the outflow-a we identify. The $v_{\rm lsr}$ of SMS1, SMS2 and SMS3 are $-39.3$~km~s$^{-1}$, $-42.8$~km~s$^{-1}$ and $-42.8$~km~s$^{-1}$. The beam of the single dish data is shown in the bottom left corner of each panel.}
\label{30m_chan}
\end{figure*}

Figure \ref{30m_chan} shows the $^{12}$CO channel map of the single dish data only. The $v_{\rm lsr}$ of SMS1, SMS2 and SMS3 are $-39.3$~km~s$^{-1}$, $-42.8$~km~s$^{-1}$ and $-42.8$~km~s$^{-1}$, respectively. A red-shifted outflow component which extends towards the southwest is found associated with SMS1. However, the blue-shifted component is relatively confined around the continuum peak. We believe this outflow is in the northeast-southwest (NE-SW) direction. In the SMS2 region, one blue-shifted emission structure towards the northwest and northeast which is associated with the SMA continuum source SMS2-MM2 was detected (from panel $-52.8$~km~s$^{-1}$ to $-45.6$~km~s$^{-1}$ in Fig.~\ref{30m_chan}). However, the single-dish channel map does not show much red-shifted emission in this region. Similarly in the SMS3 region, the single-dish channel map does not show much high velocity emission. We combine the SMA data and 30~m data to study the outflow properties.

\begin{figure*}[htbp]
   \centering
    \includegraphics[angle=0,width= 15cm]{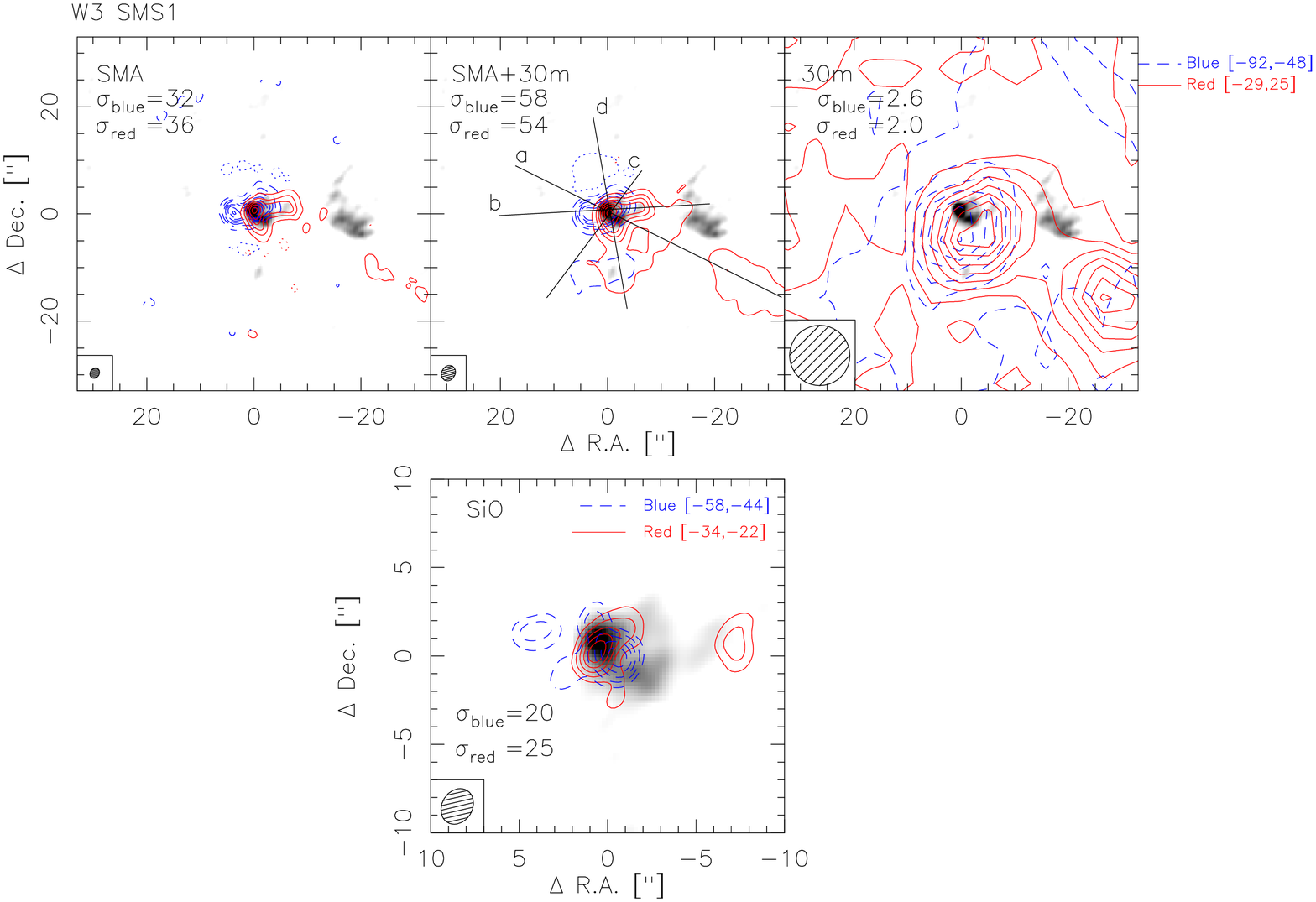}
    \caption{W3 SMS1 outflow maps observed with the SMA and IRAM 30~m telescope. The grey scale in all panels is the SMA 1.3~mm continuum image. {\it Top panels:} The full and dashed contours show the red- and blue-shifted $^{12}$CO(2$-$1) outflow emission, respectively. The top left panel presents the SMA data only map, the middle one is the combined SMA+30~m data and the right one is the IRAM 30~m data only map. The velocity-integration ranges are shown in the top right. All the CO emission contour levels start at 4$\sigma$ and continue in steps of 8$\sigma$. The four solid lines in the SMA+30m panel show the PV-cuts presented in Fig.~\ref{sms1_out_pv}. {\it Bottom panel:} The full and dashed contours show the red- and blue-shifted SiO($5-4$) outflow emission, respectively. All the SiO emission contour levels start at 4$\sigma$ and continue in steps of 2$\sigma$. The $\sigma$ value for each outflow emission is shown in the respective map in mJy~beam$^{-1}$, except for the 30~m data only map which has units of K~km~s$^{-1}$. The dotted contours are the negative features due to the missing flux with the same contour levels as the positive ones in each panel. The synthesized beam of the outflow images is shown in the bottom left corner of each panel. The (0, 0) point in each panel is R.A. 02$^{\rm h}$25$^{\rm m}$40.68$^{\rm s}$ Dec. $+62^{\circ}$05$'$51.5$''$ (J2000).}
\label{sms1_out}
\end{figure*}

\begin{figure}[htbp]
   \centering
    \includegraphics[angle=0,width=\columnwidth]{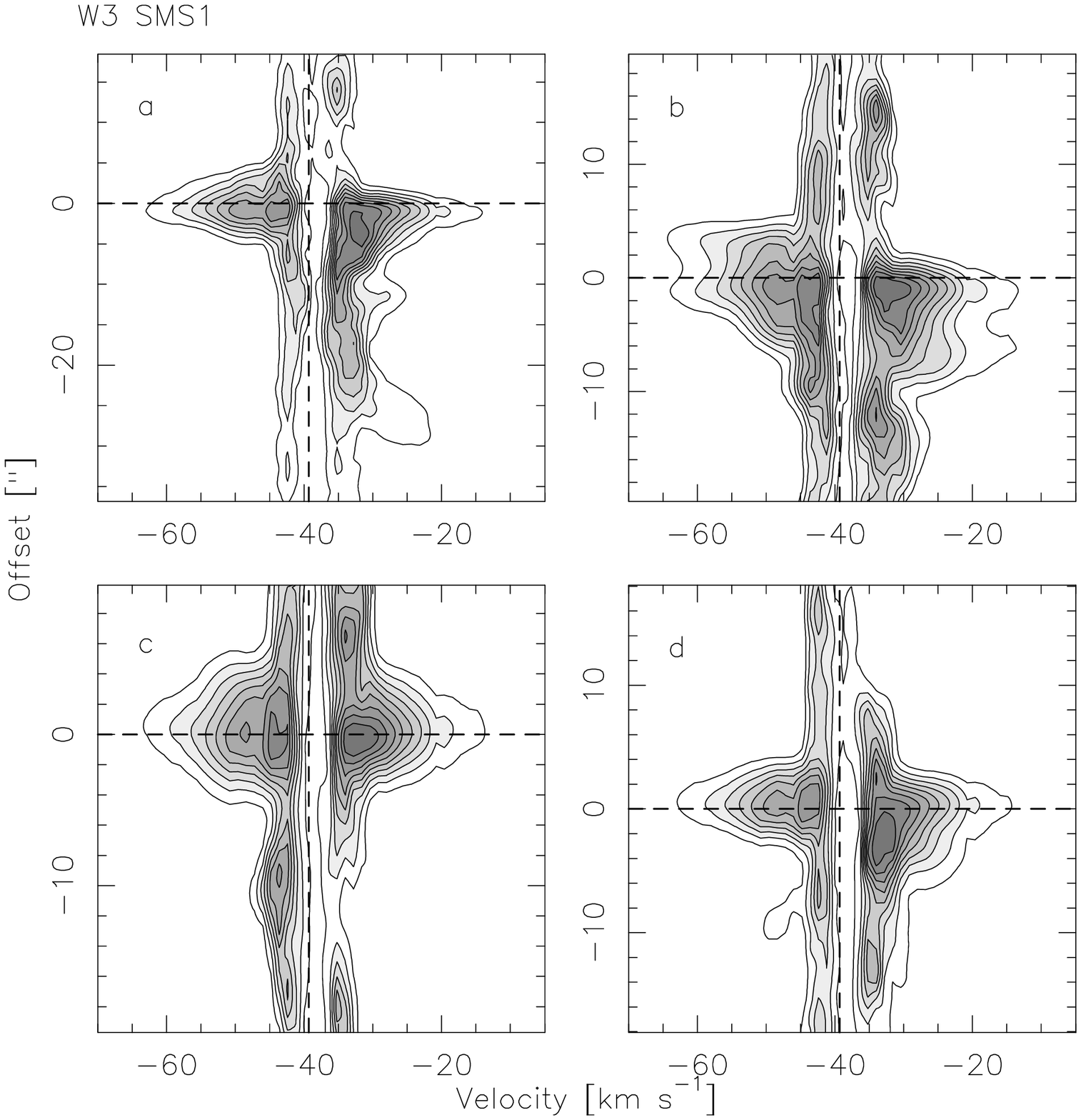}
    \caption{Position-velocity diagrams of W3 SMS1 for the $^{12}$CO SMA+30m outflow observations with a velocity resolution of 1.2~km~s$^{-1}$. The PV-cut of each panel is shown in the {\it top-middle panel} of Fig.~\ref{sms1_out} as the same letter a, b, c and d. The contour levels are from 10 to 90$\%$ from the peak emission (30.3 Jy beam$^{-1}$ for panel a, 26.6 Jy beam$^{-1}$ for panel b, 28.3 Jy beam$^{-1}$ for panel c, 31.4 Jy beam$^{-1}$ for panel d) in steps of 10$\%$. The $v_{\rm lsr}$ at $-39.3$~km~s$^{-1}$ and the central position are marked by vertical and horizontal dashed lines. The central position of panel a and b is the SMS1-MM1 source, and for panel c and d it is the red-shifted outflow peak at R.A. 02$^{\rm h}$25$^{\rm m}$40.61$^{\rm s}$ Dec. $+62^{\circ}05'51.7''$ (J2000).}
\label{sms1_out_pv}
\end{figure}

\paragraph{W3 SMS1}
The top panels in Fig.~\ref{sms1_out}, from left to right, show the $^{12}$CO($2-1$) outflow image of the W3 SMS1 region derived from the SMA data only, combined SMA+30~m data and the IRAM 30~m data only. The bottom panel in Fig.~\ref{sms1_out} shows the SiO(5$-4$) outflow map observed with SMA. \citet{imai2000} and \citet{rodon2008} observed multiple outflows which are associated with W3 IRS5, and Fig.~\ref{sms1_out} also shows the complicated outflow environment. By comparing our outflow data with previous work which has been done for this region \citep{claussen1984, mitchell1991, choi1993, hasegawa1994, imai2000, gibb2007, rodon2008}, we identify four outflows in this region and mark them with four lines in the SMA+30m panel of Fig.~\ref{sms1_out}. Figure \ref{sms1_out_pv} shows the position-velocity (PV) diagrams of the $^{12}$CO SMA+30m outflow observations. 

Outflow-a: The NE-SW outflow shown in Fig.~\ref{30m_chan} is denoted as outflow-a, which is also seen in all the $^{12}$CO outflow panels of Fig.~\ref{sms1_out}. The blue-shifted part of outflow-a is not as prominent as the red-shifted part. This outflow and its asymmetric structure has also been detected by \citet{claussen1984, mitchell1991, choi1993} and \citet{hasegawa1994}. Since the CO outflow emission traces the entrained gas \citep{arce2007}, it can be significantly affected by the environment. Therefore, the asymmetric structure of outflow-a could be due to different properties of the ambient gas. Outflow ``Flow~A'' identified with H$_2$O maser observations by \citet{imai2000} is in a similar direction as outflow-a. We believe that the outflow ``Flow~A'' and our outflow-a actually trace different parts of the same outflow. Outflow ``Flow A'' traces the inner small scale structure and outflow-a traces the outer large scale structure. The different directions between these two outflows could be due to precession of this outflow caused by the relative motions of the driving source of outflow ``Flow A'' with respect to that of ``Flow B'' found by \citet{imai2000}. However, this outflow does not show up in our SiO observations or the SiO observations done by \citet{rodon2008}. The emission of these two molecules does not necessarily stem from exactly the same environment because unlike CO outflow emission, SiO emission traces strong shocks in dense molecular gas \citep{schilke1997}. The other three outflows can only be detected in the SMA and SMA+30m panel. The PV diagram of outflow-a (panel a, Fig.~\ref{sms1_out_pv}) shows that the far end of the red-shifted part resembles the outflow Hubble-law with increasing velocity at a larger distance from the outflow center (e.g., \citealt{arce2007}).

Outflow-b: Outflow-b is more or less in the east-west (E-W) direction, and is also shown in the SiO outflow map. The PV diagram (panel b, Fig.~\ref{sms1_out_pv}) shows a weak Hubble-law profile \citep{arce2007}. This outflow was first detected by SiO emission \citep{gibb2007} and is only detected in the better resolution SMA and SMA+30m maps.  \citet{rodon2008} claimed that one of their SiO outflows (their SiO-c) could represent the SiO outflow detected by \citet{gibb2007} (i.e. our outflow-b). However, the outflow detected by \citet{rodon2008} has the red-shifted part in the east and the blue-shifted part in the west, which is exactly the opposite of our outflow-b. Thus, we believe this SiO outflow  traces a different outflow than the one in \citet{rodon2008}.

Outflow-c: All the PV diagrams (Fig.~\ref{sms1_out_pv}) are dominated by one strong outflow which is aligned very close to the line of sight at or near the zero-offset point. \citet{rodon2008} also reported an outflow which is aligned very close to the line of sight. Along the line c in Fig.~\ref{sms1_out}, the SiO outflow map shows an opposite red-blue direction compared to the CO outflow map in the vicinity of SMS1-MM1. Thus we suggest there is another outflow, outflow-c, which is aligned close to the line of sight. Line c in Fig.~\ref{sms1_out} does not represent the direction of outflow-c, we just choose it to produce the PV diagram (panel c in Fig.~\ref{sms1_out_pv}), which shows a symmetric velocity profile. This outflow is also detected in the SiO outflow map (Fig.~\ref{sms1_out}). \citet{rodon2008} identified two SiO outflows which are aligned close to the line of sight, and outflow-c is likely the joint contribution of these two SiO outflows. Figure \ref{sms1_so} shows that the emission from all the SO$_2$ and its isotopologues peaks on the SMA continuum source SMS1-MM1. Since SO$_2$ is known to be enriched by shock interactions with outflows \citep{beuther2009}, the strong SO$_2$ emission at all different energy levels could be due to the shock driven by outflow-c. 

Outflow-d: Outflow-d is roughly in the north-south direction, and is also shown in the SiO outflow map. This N-S outflow matches the water maser outflow found by \citet{imai2000} and one of the outflows detected from proper motions of compact radio sources by \citet{wilson2003}. 

The bottom panel of Fig.~\ref{sms1_out} shows that the central part of the blue-shifted SiO outflow emission is offset to the SW direction by $\sim2''$ from the continuum peak of SMS1-MM1. This offset could be produced by the outflow SiO-c identified by \citet{rodon2008} (the red-shifted part is in the E direction and the blue-shifted part is in the W direction). Two additional outflows identified by \citet{rodon2008} (their SiO-d and SiO-e) are not found in our observations. In principle, we cannot exclude that this outflow is just a wide-angle outflow, however, from our point of view the diverse data and the observations mentioned above rather favour a multiple outflow scenario.

The outflow map of SMS1 shows that the driving source of all the outflows we found here is the SMA continuum source SMS1-MM1, which was resolved into four 1.4 mm continuum sources by \citet{rodon2008}. They claim the driving source of their outflow SiO-e may be the NH$_3$ feature close to the projected center of SiO-e. This NH$_3$ feature is associated with our SMA continuum source SMS1-MM4, and we do not see an outflow driven by this source. The existence of multiple outflows driven by SMS1-MM1 also confirms that SMS1-MM1 is a proto-Trapezium system as proposed by \citet{megeath2005}.

\begin{figure*}[htbp]
   \centering
    \includegraphics[angle=0,width= 15cm]{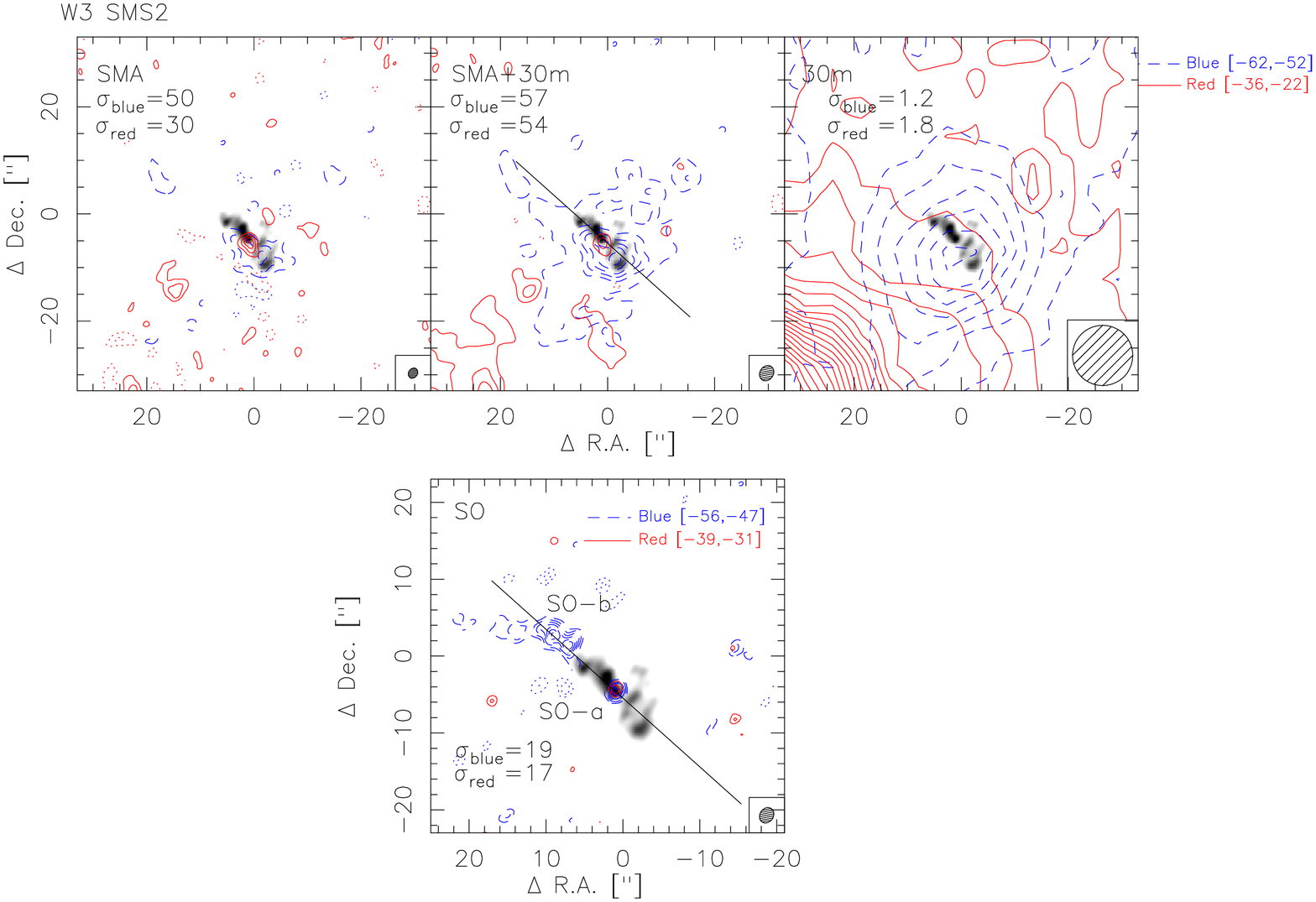}
    \caption{W3 SMS2 outflow images observed with the SMA and IRAM 30~m telescope. The grey scale in all panels is the SMA 1.3~mm continuum image. {\it Top panels:} The full and dashed contours show the red- and blue-shifted $^{12}$CO(2$-$1) outflow emission. The top left panel presents the SMA data only map, the middle one is the combined SMA+30~m data and the right one is the IRAM 30~m data only map. The velocity-integration regimes are shown in the top right. All the $^{12}$CO emission contour levels start at 4$\sigma$, for the blue-shifted outflow the contour step is 8$\sigma$ and for the red-shifted outflow it is 4$\sigma$. {\it Bottom panel:} The full and dashed contours show the red- and blue-shifted SO($6_5-5_4$) outflow emission. All the SO emission contour levels start at 4$\sigma$  and continue in steps of 2$\sigma$. The solid lines in the SMA+30m panel and the SO panel show the PV-cuts presented in Fig.~\ref{sms2_out_pv}. The $\sigma$ value for each outflow emission is shown in the respective map in mJy~beam$^{-1}$, except for the 30~m data only map which has units of K~km~s$^{-1}$. The dotted contours are the negative features due to the missing flux with the same contour levels as the positive ones in each panel. The synthesized beam of the outflow images is shown in the bottom right corner of each panel. The (0, 0) point in each panel is R.A. 02$^{\rm h}$25$^{\rm m}$31.22$^{\rm s}$ Dec. $+62^{\circ}06'25.52''$ (J2000).}
\label{sms2_out}
\end{figure*}

\paragraph{W3 SMS2}
The top panels in Fig.~\ref{sms2_out} show the W3 SMS2 $^{12}$CO($2-1$) outflow maps. The SMA and SMA+30m panels show that the blue-shifted outflow emission is strong and extended. Although the red-shifted outflow emission is relatively weak and confined to the SMA continuum source SMS2-MM2 region, it still has a peak intensity of 16$\sigma$ (1$\sigma=$ 30 mJy beam$^{-1}$) in the SMA outflow map. In the 30~m only outflow map, the red-shifted outflow emission is smoothed out and only the strong blue-shifted emission remains. 

The bottom panel in Fig.~\ref{sms2_out} shows the SO($6_5-5_4$) outflow map observed with the SMA. The red-shifted SO emission coincides with the blue-shifted component SO-a and only peaks on SMS2-MM2, the same as the $^{12}$CO($2-1$) red-shifted emission. The blue-shifted SO emission shows two lobes, one lobe peaks on the continuum source SMS2-MM2 that we denote as SO-a, and the other one extends towards the northeast direction that we denote as SO-b. There is also blue-shifted $^{12}$CO emission in the SO-b direction, but it is not as clear as SO-b. Since the UCH{\scriptsize II} region W3 C is next to SMS2-MM2, the ambient molecular gas of SMS2-MM2 could easily be blown away and affect the outflow by making the red- and blue-shifted outflow components asymmetric. We believe that SMS2-MM2 is the driving source of this outflow and the outflow is in the northeast-southwest direction. However, we cannot rule out the possibility of multiple outflows.

\begin{figure}[htbp]
   \centering
    \includegraphics[angle=0,width=\columnwidth]{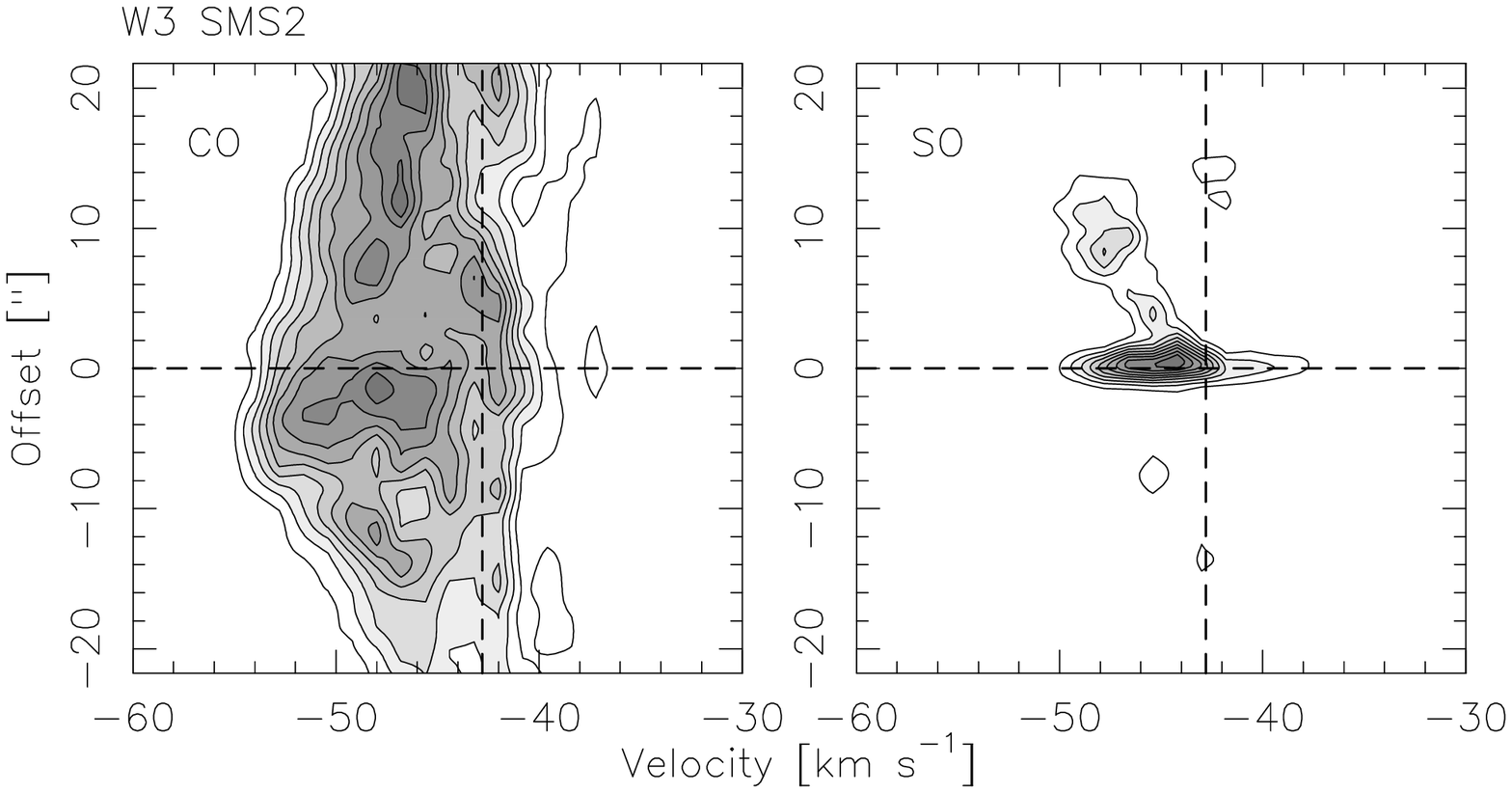}
    \caption{Position-velocity diagram of W3 SMS2 for the $^{12}$CO(2$-1$) SMA+30m data and the SO data observations with a velocity resolution of 1.2~km~s$^{-1}$. The PV-cut is in the NE-SW direction with a PA of $\sim50^{\circ}$ E of N (the cut is shown in Fig.~\ref{sms2_out}). The contour levels are from 10 to 90$\%$ from the peak emission (16.9 Jy beam$^{-1}$ for the CO panel and 2.6 Jy beam$^{-1}$ for the SO panel) in steps of 10$\%$. The $v_{\rm lsr}$ at $-42.8$~km~s$^{-1}$ and the central position (i.e. the SMS2-MM2 peak position) are marked by vertical and horizontal dashed lines.}
\label{sms2_out_pv}
\end{figure}

The left panel of Fig.~\ref{sms2_out_pv} shows the PV diagrams of the combined SMA+30m $^{12}$CO$\mbox(2-1)$ observations. The PV-plot cuts follow the straight lines shown in Fig.~\ref{sms2_out}, top middle panel. The diagram is dominated by strong blue-shifted emission, and the red-shifted emission is quite weak. The right panel of Fig.~\ref{sms2_out_pv} shows the PV diagram of the SMA SO($6_5-5_4$) observations. The PV-plot cuts follow the straight lines shown in Fig.~\ref{sms2_out}, bottom panel. The diagram shows that a pair of blue- and red-shifted emission features remain close to the center of the outflow (i.e. the SMS2-MM2 peak position) while another blue-shifted lobe resembles the Hubble-law with increasing velocity at a larger distance from the outflow center \citep{arce2007}.

\citet{ridge2001} found a bipolar outflow associated with W3 SMS1 in the direction NW-SE. The red-shifted emission of this outflow peaks towards SMS1 but the blue-shifted emission extends towards SMS2. We do not detect this outflow, however, we do detect outflows in SMS2 (Fig.~\ref{sms2_out}). Although the red-shifted emission is not as strong as the blue-shifted emission, it is prominent enough to show that the outflow is associated with SMS2-MM2. We believe that the blue-shifted outflow emission \citet{ridge2001} detected is driven by SMS2-MM2 not SMS1-MM1 (aka. W3 IRS5). Furthermore, there is a $\sim4$~km~s$^{-1}$ difference between SMS1 and SMS2. One could easily mistake the CO emission at the $v_{\rm lsr}$ of SMS2 as blue-shifted outflow emission.

\begin{figure*}[htbp]
   \centering
    \includegraphics[angle=0,width= 15cm]{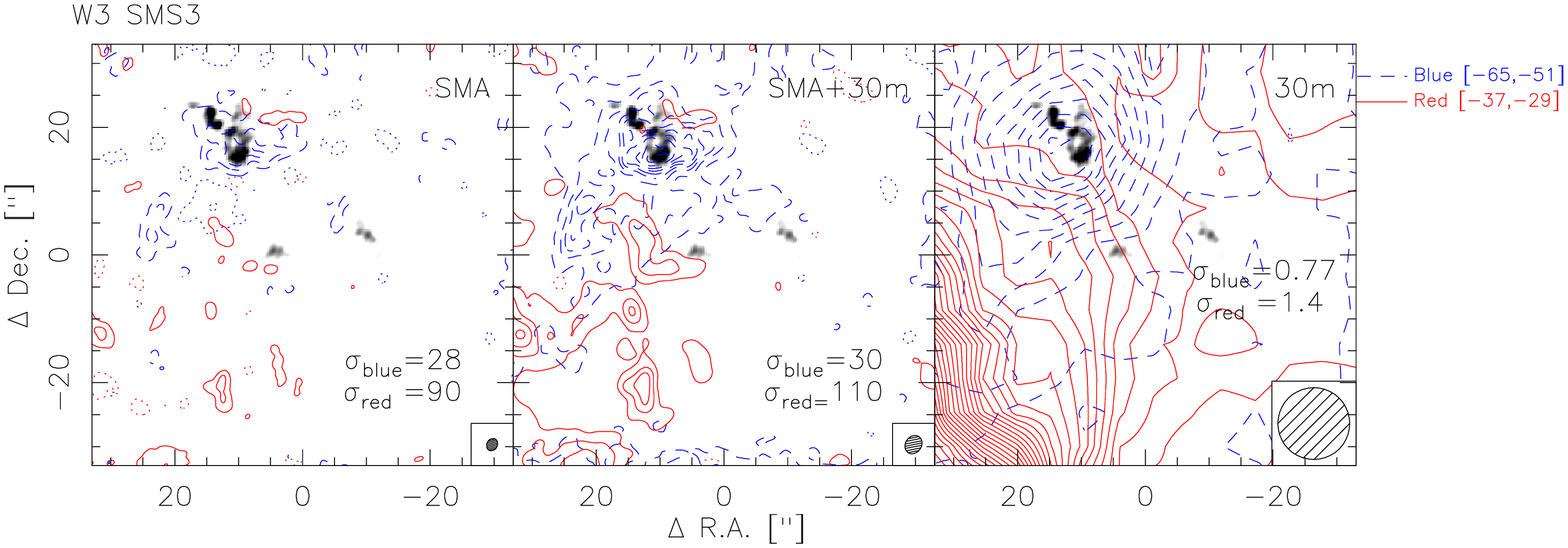}
    \caption{W3 SMS3 outflow images observed with the SMA and IRAM 30~m telescope. The grey scale in all panels is the SMA 1.3~mm continuum image. The full and dashed contours show the red- and blue-shifted $^{12}$CO(2$-$1) outflow emission.  The top left panel presents the SMA data only map, the middle one is the combined SMA+30~m data and the right one is the IRAM 30~m data only map. The velocity-integration regimes are shown in the top right. All the $^{12}$CO emission contour levels start at 4$\sigma$, for the blue-shifted outflow the contour step is 8$\sigma$ and for the red-shifted outflow it is 4$\sigma$. The $\sigma$ value for each outflow  is shown in the respective map in mJy~beam$^{-1}$, except for the 30~m data only map which has units of K~km~s$^{-1}$. The dotted contours are the negative features due to the missing flux with the same contour levels as the positive ones in each panel. The synthesized beam of the outflow images is shown in the bottom right corner of each panel. The (0, 0) point 
in each panel is R.A. 02$^{\rm h}$25$^{\rm m}$29.49$^{\rm s}$ Dec. $+62^{\circ}06'00.6''$ (J2000).}
\label{sms3_out}
\end{figure*}

\paragraph{W3 SMS3}
Fig.~\ref{sms3_out} shows the combined SMA and 30~m outflow images of W3 SMS3. All panels exhibit strong blue-shifted emission associated with SMS2. Some weak red- and blue-shifted features are detected toward SMS3, but we do not find any outflow signature associated with SMS3. Perhaps the outflow activity is weak at SMS3, and due to the interference from the strong emission at SMS2, we cannot recover the outflow emission. This again indicates that SMS3 is at an early evolutionary stage, with hardly any star formation activity.

\subsection{Rotational structures}
We use SO$_2(22_{7,15}-23_{6,16})$, H$_2$CO$\mbox(3_{0,3}-2_{0,2})$, H30$\alpha$, CH$_3$OH$\mbox(4_{2,2}-3_{1,2})$E and HC$_3$N$\mbox(24-23)$ to study the rotational properties of the three regions. We find rotation structures in SMS1 and SMS2, and do not find any rotational signatures in SMS3.
 
\begin{figure*}[htbp]
   \centering
    \includegraphics[angle=0,width= 14cm]{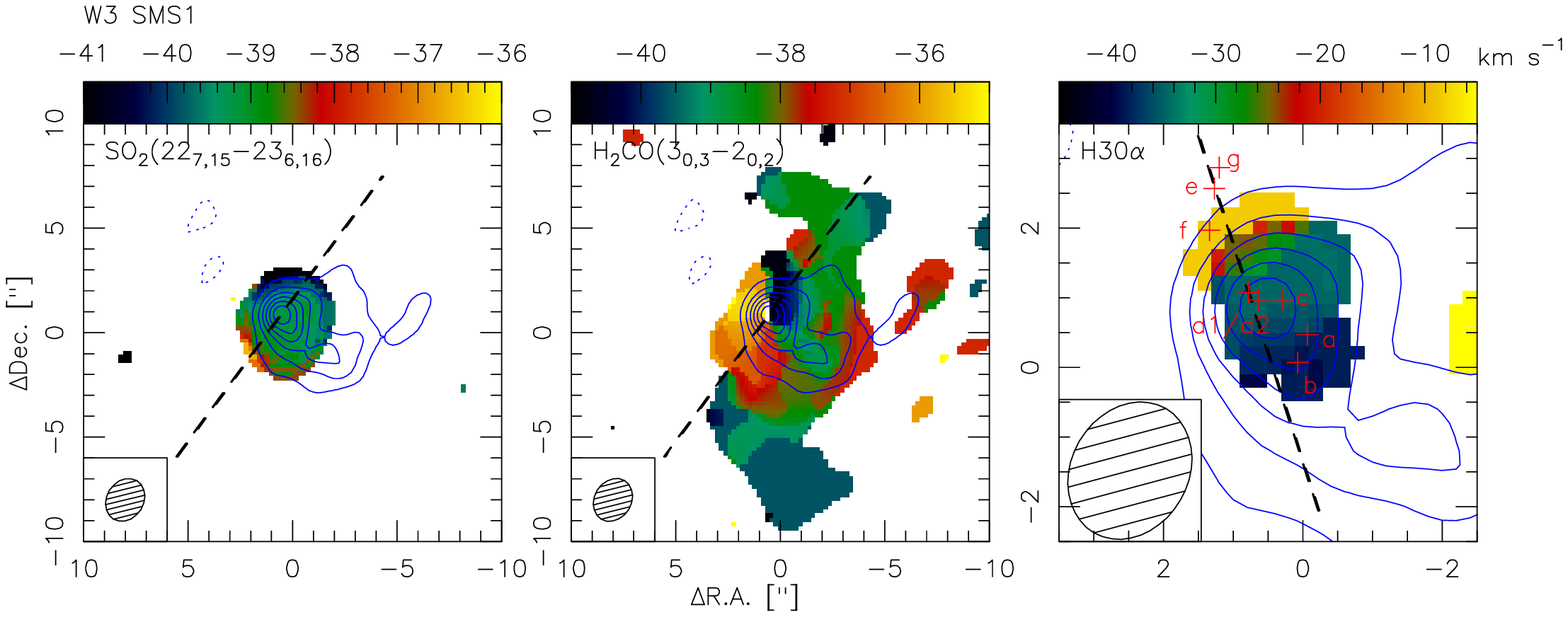}
    \caption{SO$_2(22_{7,15}-23_{6,16})$, H$_2$CO($3_{0,3}-2_{0,2}$) and H30$\alpha$ velocity (1st) moment maps overlaid with the SMA 1.3~mm dust continuum of W3 SMS1. The contours start at 5$\sigma$ and increase in steps of 5$\sigma$ in all panels (1$\sigma$=3.6~mJy~beam$^{-1}$). The dotted contours are the negative features due to the missing flux with the same contour levels as the positive ones in each panel. The dashed lines in each panel show the PV diagram cuts presented in Fig.~\ref{sms1_mom1_pv}. The crosses and the letters mark the compact radio sources detected by \citet{claussen1994} and \citet{tieftrunk1997}. All moment maps are clipped at the five sigma level of the respective line channel map. The synthesized beams of the moment maps are shown in the bottom left corner of each plot. The (0, 0) point in each panel is R.A. $02^{\rm h}25^{\rm m}40.68^{\rm s}$ Dec. $+62^{\circ}05'51.5''$ (J2000).}
\label{sms1_mom1}
\end{figure*}

\paragraph{W3 SMS1}
In the velocity map of SO$_2(22_{7,15}-23_{6,16})$ (Fig.~\ref{sms1_mom1}), we see a clear velocity gradient in the northwest-southeast (NW-SE) direction which is perpendicular to the direction of outflow-a (Fig.~\ref{sms1_out}) indicating that the envelope of SMS1-MM1 may rotate in this direction. \citet{rodon2008} also observed a similar velocity gradient from a lower sulfur dioxide transition. The H$_2$CO($3_{0,3}-2_{0,2}$) velocity map (Fig.~\ref{sms1_mom1}) shows a larger structure with complicated velocity signatures. 

The H30$\alpha$ velocity map (Fig.~\ref{sms1_mom1}) shows a velocity gradient in the NE-SW direction and follows the distribution of the compact radio sources detected by \citet{tieftrunk1997}, which is nearly perpendicular to the direction of outflow-b. Hence the H30$\alpha$ emission could be tracing the rotation of the ionized gas, although we cannot exclude the possibility of an ionized jet. This velocity gradient is different from the one seen in SO$_2(22_{7,15}-23_{6,16})$, and outflow-d is found in the similar direction but with an inverse red-blue signature (Fig.~\ref{sms1_out}). 

\begin{figure*}[htbp]
   \centering
    \includegraphics[angle=0,width=14cm]{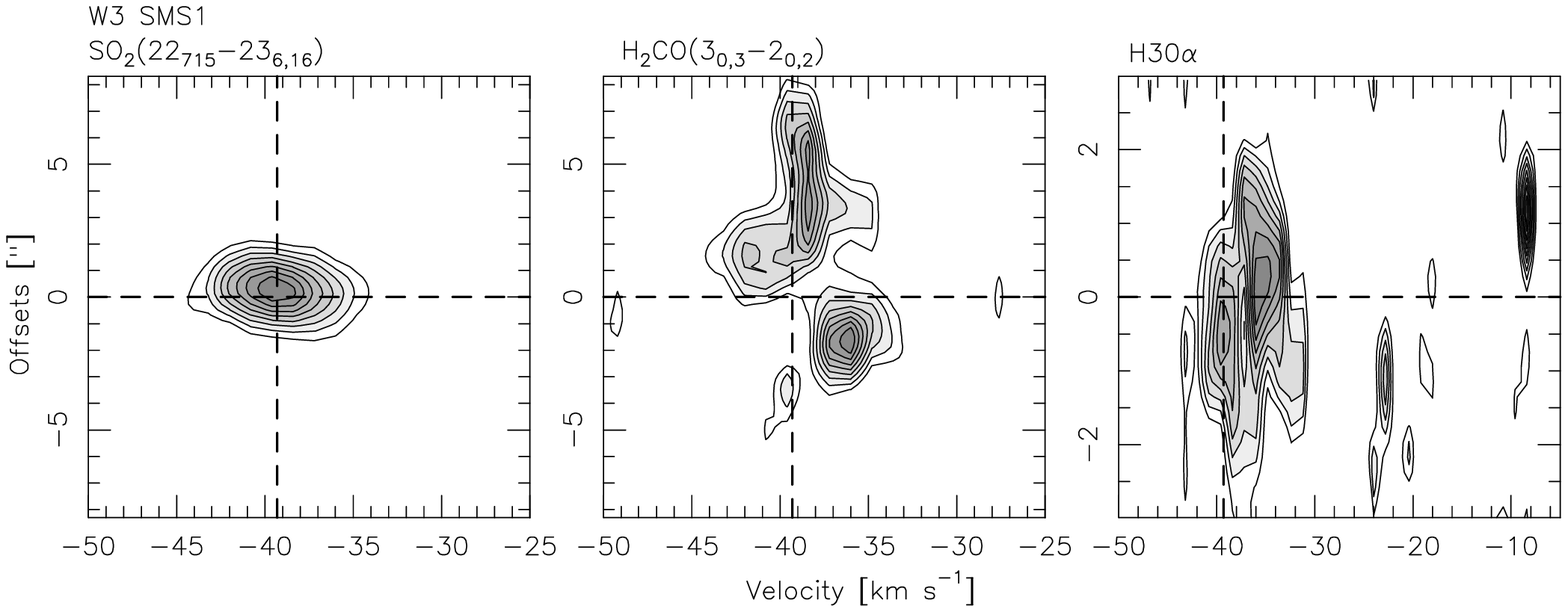}
    \caption{Position-velocity diagrams derived for the cuts along the observed velocity gradient in Fig.~\ref{sms1_mom1} with a velocity resolution of 1.2~km~s$^{-1}$. The PV-cut for H30$\alpha$ and SO$_2(22_{7,15}-23_{6,16})$ is marked in the corresponding 1st moment maps (Fig.~\ref{sms1_mom1} ). The PV-cuts for H$_2$CO($3_{0,3}-2_{0,2}$) a and b are shown as line a and b in the H$_2$CO($3_{0,3}-2_{0,2}$) velocity map (Fig.~\ref{sms1_mom1}). All the contour levels are from 10 to 90$\%$ of the peak emission (264~mJy~beam$^{-1}$ for H30$\alpha$, 833~mJy~beam$^{-1}$ for H$_2$CO($3_{0,3}-2_{0,2}$) panel a, 525~mJy~beam$^{-1}$ H$_2$CO($3_{0,3}-2_{0,2}$) panel b, 2.68 Jy beam$^{-1}$ for SO$_2(22_{7,15}-23_{6,16})$) in steps of 10$\%$. The offsets refer to the distance along the cuts from the dust continuum peak of SMS1-MM1. The $v_{\rm lsr}$ at $-39.3$~km~s$^{-1}$ and the central position (the SMS1-MM1 continuum peak) are marked by vertical and horizontal dashed lines.} 
\label{sms1_mom1_pv}
\end{figure*}

The position-velocity diagrams of the SO$_2$\mbox{$(22_{7,15}-23_{6,16})$}, H$_2$CO\mbox{$(3_{0,3}-2_{0,2})$} and H30$\alpha$ emission are shown in Fig.~\ref{sms1_mom1_pv}. The corresponding PV cuts of each panel are shown in Fig.~\ref{sms1_mom1}. The PV diagrams of SO$_2\mbox(22_{7,15}-23_{6,16})$ and H$_2$CO($3_{0,3}-2_{0,2}$) show that the rotational structure is not in Keplerian motion, hence it may be just a rotating and infalling core similar to the toroids or envelope described by \citet{cesaroni2007}. The rotational envelope detected by our SO$_2(22_{7,15}-23_{6,16})$ observation has a size of $\sim10\time10^4$~AU at the given distance of 1.95~kpc. This indicates that the proto-Trapezium system shares one common rotating envelope. The velocity jump found by \citet{rodon2008} in SO$_2$ emission is also confirmed by our H$_2$CO observations.

The PV diagram of the H30$\alpha$ emission shows quite a small velocity gradient for the radio recombination line. We extract the spectrum towards the emission peak in Table \ref{line_pos}, and derive the FWHM line width from the Gaussian fitting as $\sim$ 9.2 km s$^{-1}$. For a typical HCH{\scriptsize II} region with a temperature of 8~000 K, the line width due to thermal broadening would be 19.1~km s$^{-1}$. Hence, the H30$\alpha$ emission is likely tracing the gas associated with kinematic properties, e.g., rotation and/or outflow, but we cannot properly differentiate among the options.

\begin{figure}[htbp]
   \centering
    \includegraphics[angle=0,width=\columnwidth]{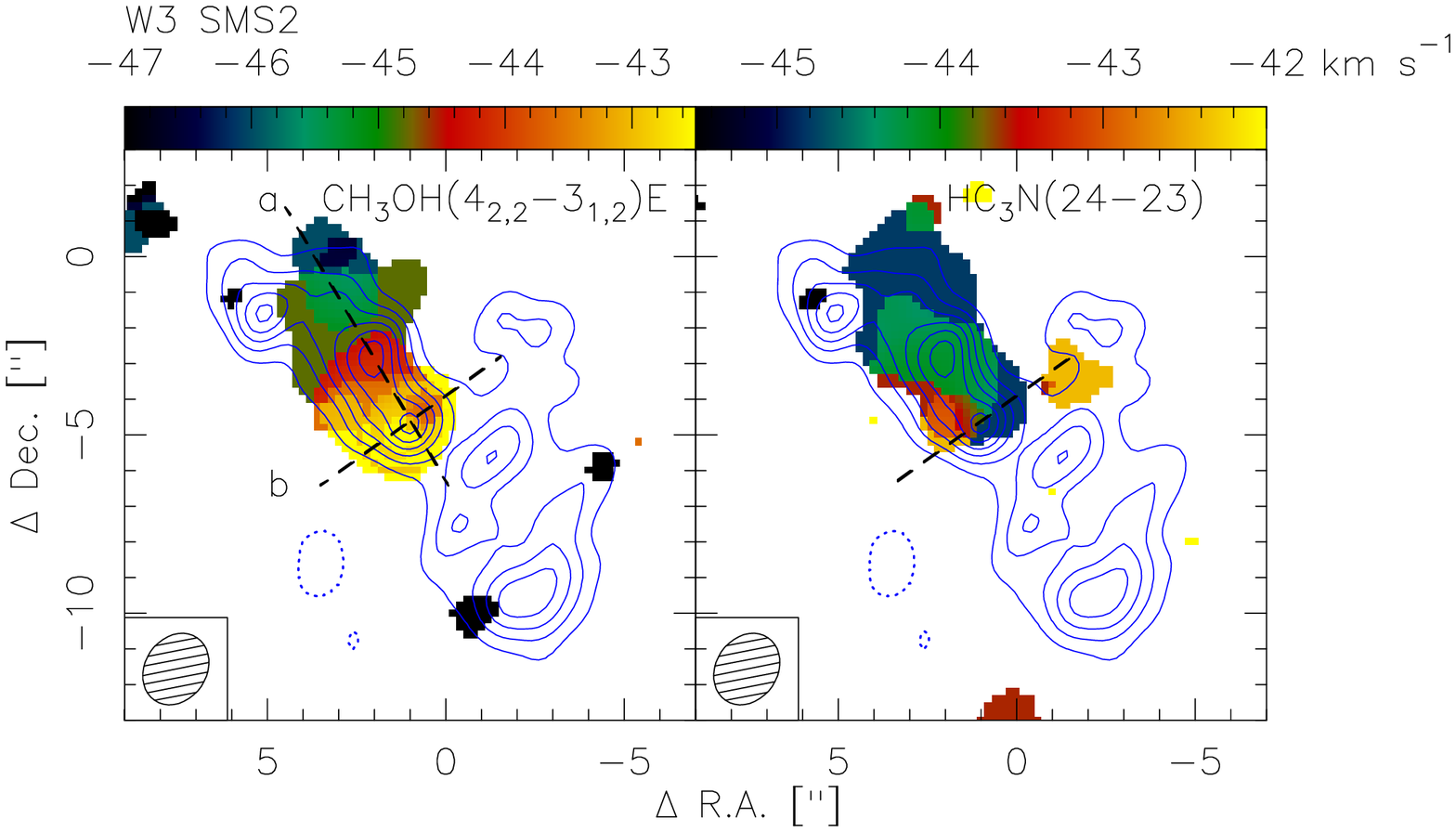}
    \caption{CH$_3$OH($4_{2,2}-3_{1,2}$)E and HC$_3$N($24-23$) velocity (1st) moment maps overlaid with the SMA 1.3~mm dust continuum of W3 SMS2. The contours start at 5$\sigma$ and increase in steps of 5$\sigma$ in all panels ($\sigma$= 2.5~mJy~beam$^{-1}$). The dotted contours are the negative features due to the missing flux with the same contour levels as the positive ones in each panel. The dashed lines in each panel show the PV diagram cuts presented in Fig.~\ref{sms2_mom1_pv}. All moment maps were clipped at the 5$\sigma$ level of the respective line channel map. The synthesized beams of the moment maps are shown in the bottom left corner of each plot. The (0, 0) point in each panel is R.A. 02$^{\rm h}$25$^{\rm m}$31.22$^{\rm s}$ Dec. $+62^{\circ}06'25.52''$ (J2000).}
\label{sms2_mom1}
\end{figure}

\paragraph{W3 SMS2}
In the velocity map of CH$_3$OH\mbox{$(4_{2,2}-3_{1,2})$}E (Fig.~\ref{sms2_mom1}), we see a clear velocity gradient in the NE-SW direction, which is in a similar direction as the blue-shifted outflow SO-b (bottom panel, Fig.~\ref{sms2_out}). CH$_3$OH is well known as a molecule that traces shocks, cores and masers in star forming regions \citep{beuther2005b, jorgensen2004, sobolev2007}, therefore, we propose that this velocity gradient traces the near-side of outflow SO-b. Another velocity gradient that is in the NW-SE direction is shown in both the CH$_3$OH$\mbox(4_{2,2}-3_{1,2})$E and HC$_3$N($24-23$) velocity maps (Fig.~\ref{sms2_mom1}). This velocity gradient is also perpendicular to the direction of outflow SO-b (bottom panel, Fig.~\ref{sms2_out}). CH$_3$OH and HC$_3$N are well known low-mass disk tracers (e.g. \citealt{goldsmith1999}), hence the NW-SE velocity gradient suggests a rotational structure.

\begin{figure*}[htbp]
   \centering
    \includegraphics[angle=0,width=14cm]{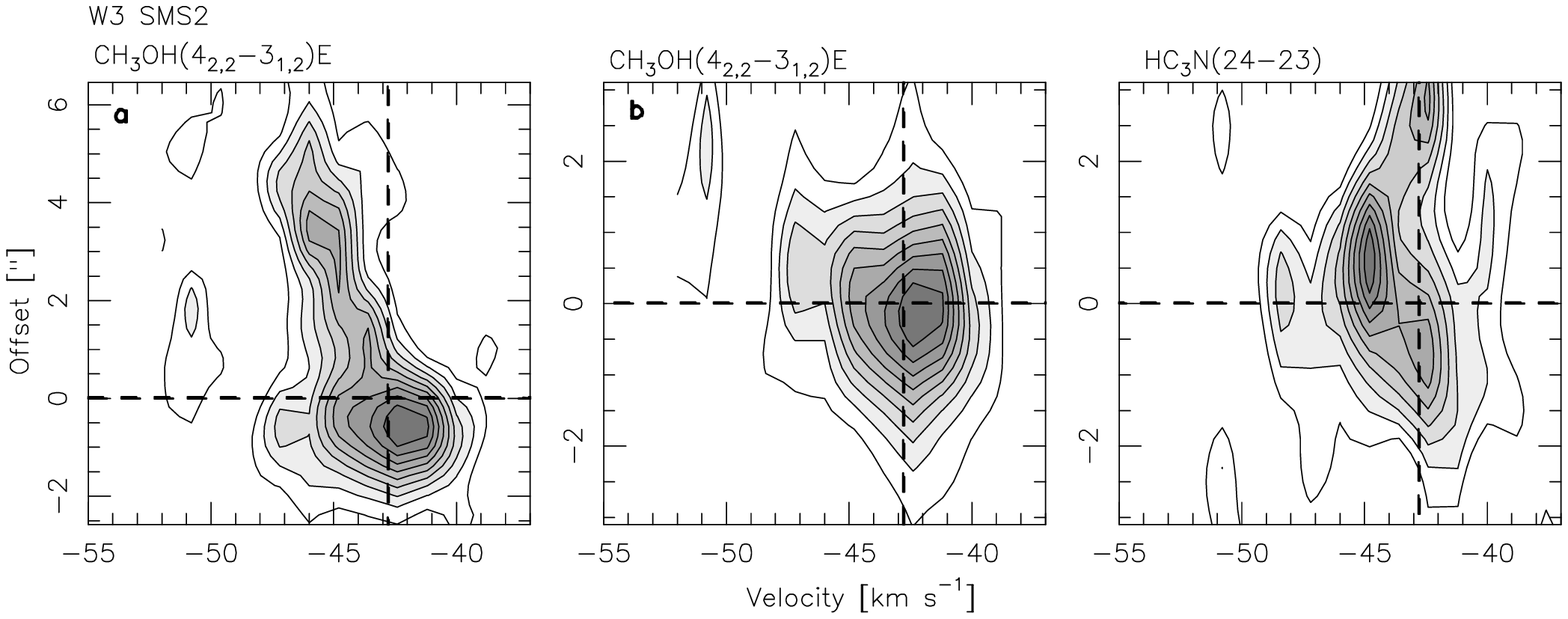}
    \caption{Position-velocity diagrams derived for the cuts along the observed velocity gradient in Fig.~\ref{sms2_mom1} with a velocity resolution of 1.2~km~s$^{-1}$. The PV-cuts for the left and middle panel are marked as line a and b in the CH$_3$OH($4_{2,2}-3_{1,2}$)E velocity map, respectively (Fig.~\ref{sms1_mom1}). The PV-cut for the HC$_3$N($24-23$) is marked as the straight line in its 1st moment map (Fig.~\ref{sms2_mom1} ). All the contour levels are from 10 to 90$\%$ of the peak emission (from left to right, 621, 644 and 370~mJy~beam$^{-1}$, respectively) in steps of 10$\%$. The offsets refers to the distance along the cuts from the dust continuum peak of SMS2-MM2. The $v_{\rm lsr}$ at $-42.8$~km~s$^{-1}$ and the central position (i.e. the SMS2-MM2 peak position) are marked by vertical and horizontal dashed lines.} 
\label{sms2_mom1_pv}
\end{figure*}

The PV diagrams of the CH$_3$OH$\mbox(4_{2,2}-3_{1,2})$E and HC$_3$N$\mbox(24-23)$ emission are shown in Fig.~\ref{sms2_mom1_pv}. The PV-cut of CH$_3$OH($4_{2,2}-3_{1,2}$)E goes through the SMS2-MM2 dust continuum peak and is marked with line a in Fig.~\ref{sms2_mom1}. The PV diagram shows that the blue-shifted CH$_3$OH$\mbox(4_{2,2}-3_{1,2})$E resembles the outflow Hubble-law with increasing velocity at a larger distance from the outflow center \citep{arce2007}. The PV-cuts of CH$_3$OH$\mbox(4_{2,2}-3_{1,2})$E panel b and HC$_3$N(24$-$23) are shown in Fig.~\ref{sms2_mom1} $[$line b in the CH$_3$OH($4_{2,2}-3_{1,2}$)E panel and the line in HC$_3$N(24$-$23) panel$]$ are perpendicular to line a (the direction of the outflow SO-b). These two PV diagrams show that the rotational structure is not in Keplerian motion. With a size of $\sim$6~300 AU (d=1.95~kpc), this structure is similar to the rotational structure described in \citet{wang2011} and \citet{fallscheer2009}, i.e. large scale structure with a non-Keplerian velocity gradient. We do not find rotational structure for SMS2-MM1, and no rotational signature is found in SMS3 either.

\subsection{Temperature determinations}
\label{sect_temp}
\begin{figure}[htbp]
   \centering
     \includegraphics[angle=0,width= 7.5cm]{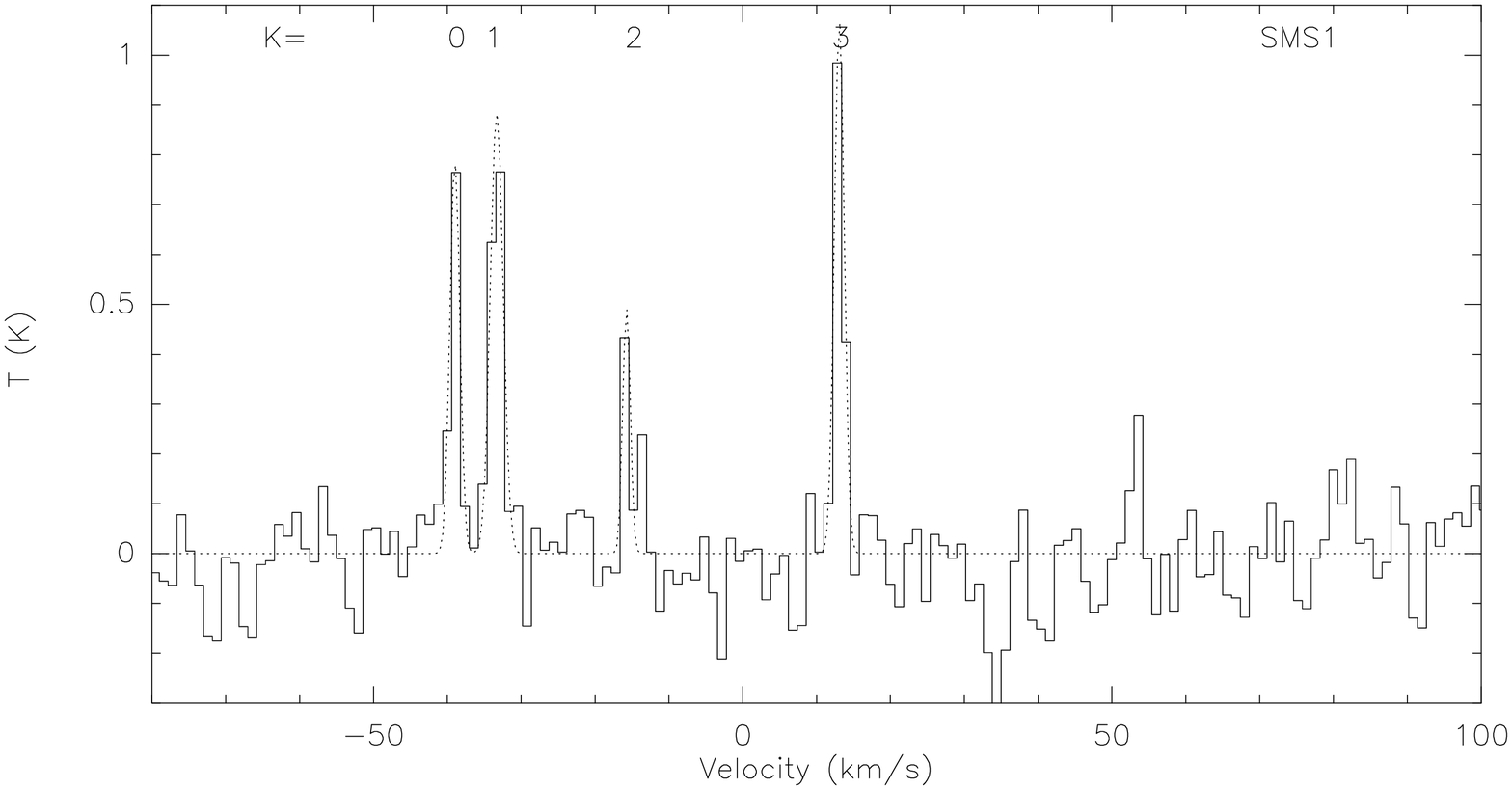}\\
     \includegraphics[angle=0,width= 7.5cm]{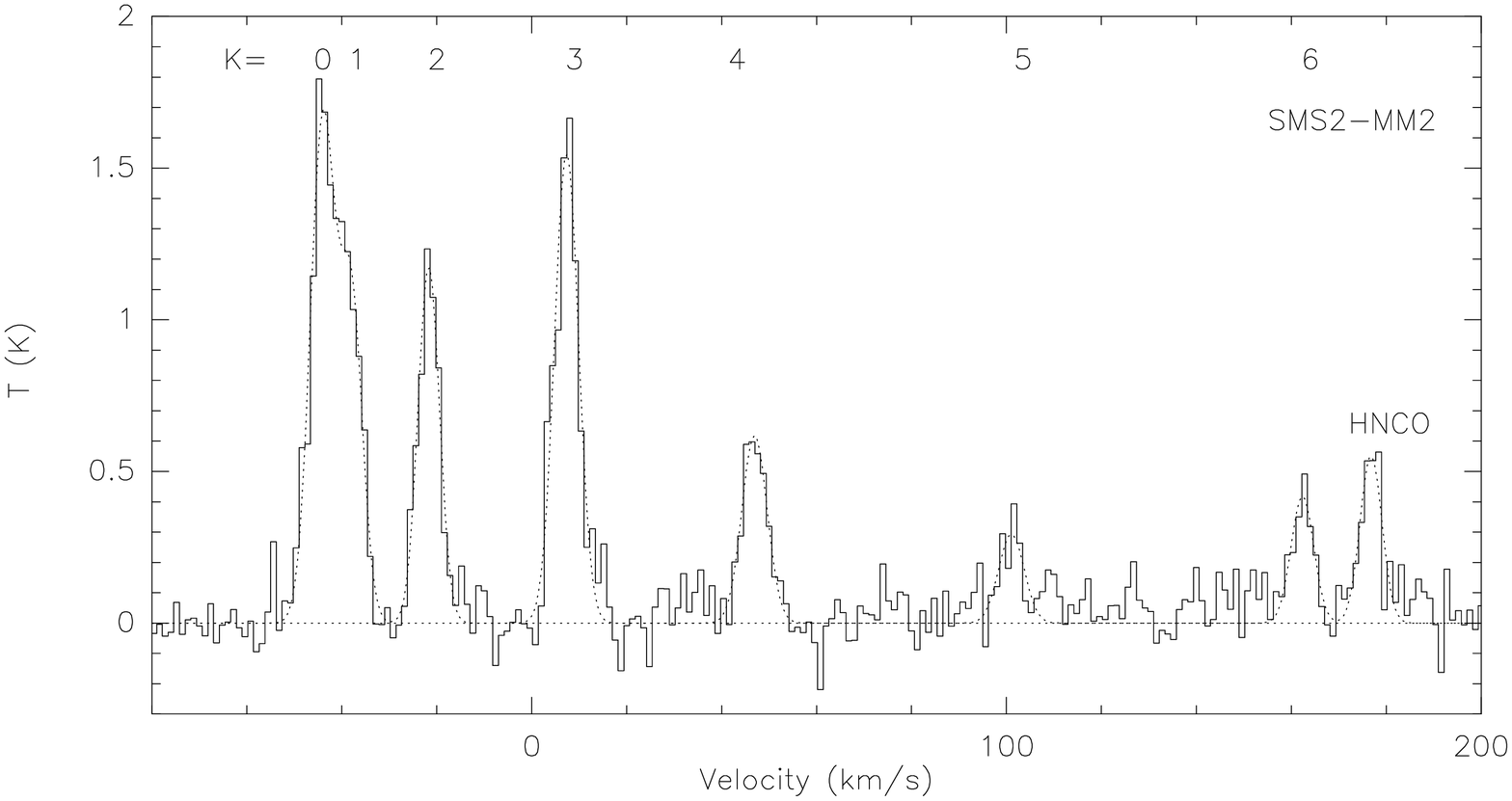}
    \caption{Observed CH$_3$CN($12_k-11_k$) spectrum toward the CH$_3$CN integrated intensity peak in SMS1 ({\it top panel}) and continuum peak SMS2-MM2 ({\it bottom panel}) with only the compact configuration data. The dotted line shows the best Gaussian fit obtained with CLASS for the lines.}
\label{ch3cn_spec}
\end{figure}

\begin{figure}[htbp]
   \centering
    \includegraphics[angle=0,width= 7.5cm]{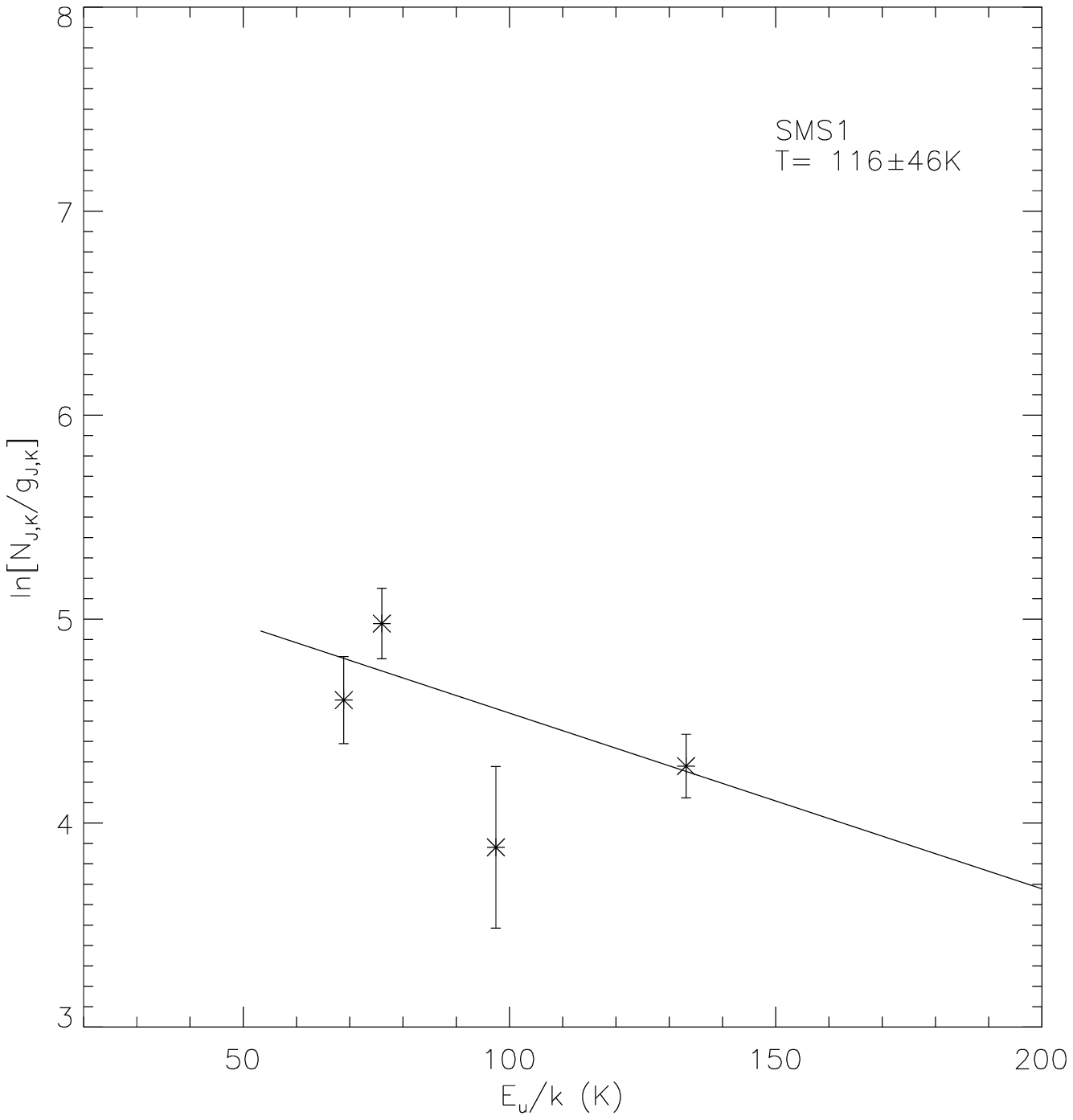}\\
    \includegraphics[angle=0,width= 7.5cm]{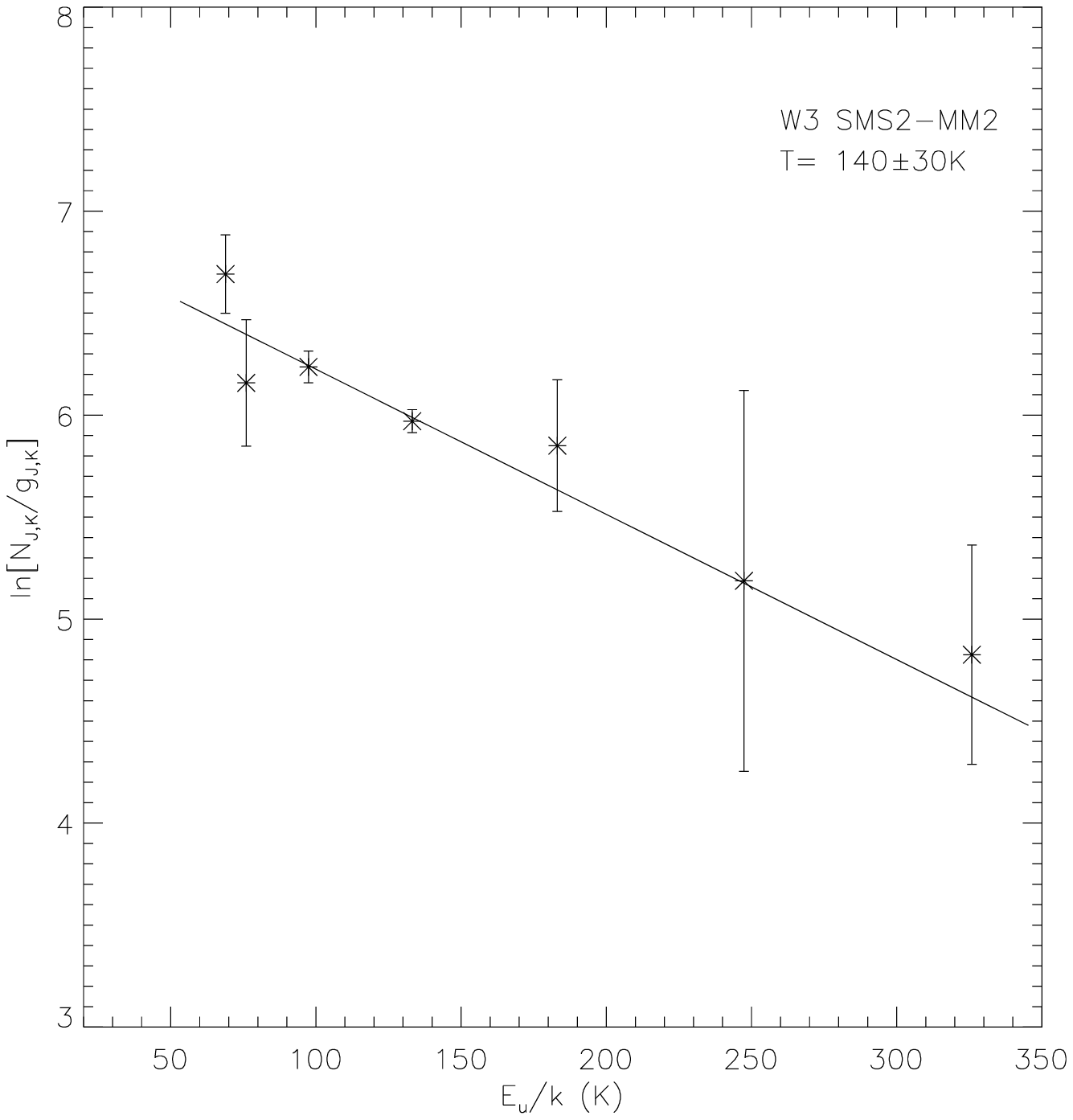}
     \caption{LTE rotational diagrams of the CH$_3$CN(12$_k-11_k$) lines shown in Fig.~\ref{ch3cn_spec}. The solid line is the linear fit to all the $k$ components calculated following the equation (B6) from \citet{zhang1998}.}
\label{ch3cnfit}
\end{figure}

The dense gas (n$\gtrsim10^5$ cm$^{-3}$) tracer CH$_3$CN can be used as a temperature determinator (e.g. \citealt{loren1984}). For W3 SMS1, four $k=0...3$ lines of the CH$_3$CN($12_k-11_k$) ladder were detected and they do not peak on any of the SMA continuum sources, so a spectrum was extracted toward the CH$_3$CN integrated emission peak with only the compact configuration data. For W3 SMS2, seven $k=0...6$ lines of the CH$_3$CN(12$_k-11_k$) ladder were detected and and they all peak on the SMA continuum source SMS2-MM2, so we extract a spectrum toward the SMS2-MM2 continuum peak with only the compact configuration data. Both spectra are shown in Fig.~\ref{ch3cn_spec}. We assume Local Thermodynamic Equilibrium (LTE) and optically thin emission, and construct the rotational diagrams following the method outlined in Appendix B of \citet{zhang1998}. The level populations $N_{j,k}$ were calculated from the Gaussian fitting results of the spectra shown in  Fig.~\ref{ch3cn_spec}. The derived rotational diagrams are shown in Fig.~\ref{ch3cnfit}, along with the linear fit of all the detected $k$ components which gives the value of T$_{rot}\sim116\pm$46 K for SMS1 and $\sim140\pm30$ K for SMS2-MM2. Studies of \citet{wilner1994} revealed that T$_{rot}$ derived with this method agrees well with those obtained from the Large Velocity Gradient (LVG) calculations. The average line-width of the CH$_3$CN($12_k-11_k$) spectra towards SMS1 (top panel, Fig.~\ref{ch3cn_spec}) is $\sim1.6$~km~s$^{-1}$. Considering our spectral resolution is 1.2~km~s$^{-1}$, the linewidth is not fully resolved. For a temperature of $\sim$116 K, the thermal line-width of CH$_3$CN is $\sim0.36$ km s$^{-1}$. For hot cores with temperatures of $\sim$100~K, the average CH$_3$CN line-width is $\gtrsim$5~km~s$^{-1}$ \citep{zhang1998,zhang2007,wang2011}. Even for some IRDCs the line-width is still $\gtrsim$2.8~km~s$^{-1}$ \citep{beuther2007a}. All these features indicate that the gas from which the CH$_3$CN emission originated is warm and at the same time shows a low level of turbulence. To our knowledge this has rarely (or not at all) been seen before. 

\begin{figure}[htbp]
   \centering
    \includegraphics[angle=0,width= 7.5cm]{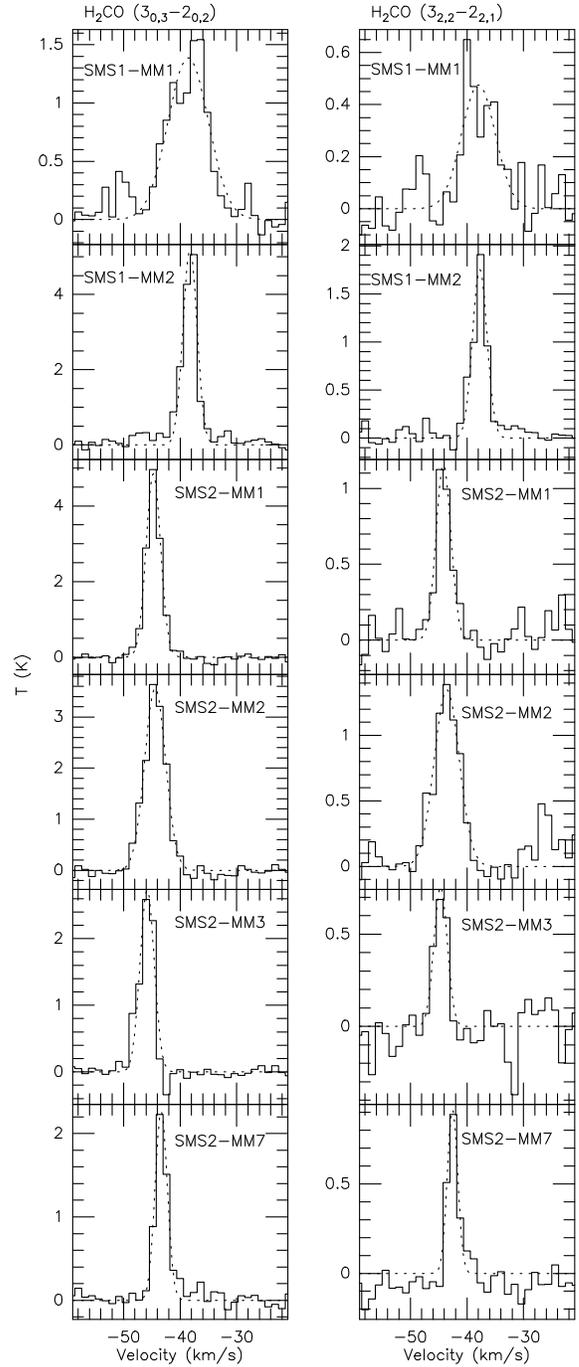}
    \caption{H$_2$CO spectra towards the SMA 1.3~mm continuum sources SMS1-MM1/2 and SMS2-MM1/2/3/7 with only the compact configuration data. The dotted lines show the Gaussian fit obtained with CLASS for each spectrum.}
\label{h2coline}
\end{figure}

H$_2$CO is a well know kinetic temperature determinator (e.g. \citealt{mangum1993, jansen1994, muehle2007, watanabe2008}). Since it is a slightly asymmetric rotor, transitions between energy levels with different $K$ are only collisionally excited. Therefore, comparing the level populations with different $K$ components from the $\Delta J=1$ transitions gives an estimate of the kinematic temperature of the medium \citep{mangum1993}. The spectra of H$_2$CO($3_{0,3}-2_{0,2}$) and H$_2$CO($3_{2,2}-2_{2,1}$) were extracted towards the continuum peaks of SMS1-MM1/2 and SMS2-MM1/2/3/7 with the compact configuration data. The spectra were processed with CLASS and are shown in Fig.~\ref{h2coline}. We assume LTE and optically thin emission, and calculate the temperature T$_{\rm kin}$ of these sources following the method described by \citet{mangum1993}. The temperature results are shown in Table \ref{temp_table}. The high temperature of SMS2-MM7 (98$\pm$33 K) could be due to the fact that it is heated by the nearby infrared source IRS4-a (Fig.~\ref{sms2_ks}).

\begin{deluxetable}{lrrc}
\tablewidth{0cm}
\tablecaption{Temperature estimation results for SMS1 and SMS2.}
\setlength{\tabcolsep}{0.1cm}
\tabletypesize{\footnotesize}
\centering
\tablehead{
\colhead{Source}&
\colhead{T$_{rot}$\tablenotemark{a}}&
\colhead{T$_{\rm kin}$\tablenotemark{b}}&
\colhead{$\int T_b({\rm H_2CO}(3_{0,3}-2_{0,2}))dv$}\\
\cline{4-4}
\colhead{}&
\colhead{(K)}&
\colhead{(K)}&
\colhead{$\int T_b({\rm H_2CO}(3_{2,2}-2_{2,1}))dv$}
}
\startdata
   SMS1-MM1 &$-$ 		& 71$\pm$18	&3.5\\
   SMS1-MM2 &116$\pm$46\tablenotemark{c} & 101$\pm$16	&2.9\\
\hline
\noalign{\smallskip}
   SMS2-MM1 &$-$ 		& 57$\pm$6	&4.1\\
   SMS2-MM2 &140$\pm$30 & 231$\pm$97	&2.2\\
   SMS2-MM3 &$-$ 		& 62$\pm$14	&3.8\\
   SMS2-MM7 &$-$ 		& 98$\pm$33	&2.9
\enddata
\tablenotetext{a}{T$_{rot}$ is obtained from the rotational diagram of the CH$_3$CN lines.}
\tablenotetext{b}{T$_{\rm kin}$ is obtained from the H$_2$CO line intensity ratios.}
\tablenotetext{c}{The spectrum we use to estimate this temperature is extracted from the CH$_3$CN emission peak, which is close to SMS1-MM2, so we put this number here.}
\label{temp_table}
\end{deluxetable}

Despite the possibility that the H$_2$CO emission is not really optically thin and it might be affected by the missing flux, the T$_{\rm kin}$ we obtain from the H$_2$CO intensity ratio agrees quite well with the T$_{rot}$ we obtain from the CH$_3$CN(12$_k-11_k$) line fit for SMS1. The CH$_3$CN and CH$_3$OH integrated intensity maps (Fig.~\ref{sms1_others}) indicate that SMS1-MM2 has a higher temperature than SMS1-MM1, and this is supported by the T$_{\rm kin}$ we obtain from the H$_2$CO intensity ratio.

The LVG model calculations done by \citet{mangum1993} show that temperatures determined from the H$_2$CO($3_{0,3}-2_{0,2}$) and H$_2$CO$\mbox(3_{2,2}-2_{2,1})$ line intensity ratios have a large dispersion depending on the volume density where the line ratio is less than 3.5, which corresponds to a temperature of 70 K. Since we do not have the volume density information for our continuum sources, the temperature determination can only be done between 20 and 70 K for line ratios larger than 3.5. Therefore, the T$_{\rm kin}$ we obtain for SMS2-MM2 is not reliable. Nevertheless, the temperature we obtain from the CH$_3$CN and H$_2$CO measurements confirms the hot molecular core (HMC) nature of the continuum source SMS2-MM2.

\section{Discussion}
\subsection{Different evolutionary stages of the continuum sources}
The SMA and IRAM 30~m data together reveal three massive star formation regions in different evolutionary stages. With our high resolution SMA observations ($\sim$2~900~AU at the given distance of 1.95 kpc) we find that all SCUBA continuum sources fragment into several cores. 

W3 SMS3 is considered to be the youngest region since no radio or infrared sources are found here \citep{megeath2008}. Our results also support this, in particular the low number of molecular lines and the absence of outflow emission. The large amount of continuum flux filtered out in our observations indicates that the gas in this region is relatively uniformly distributed. VLA observations of NH$_3$(1,1) and NH$_3$(2,2) found extended emission toward SMS3, and furthermore, this extended emission was resolved into four NH$_3$ clumps surrounding the core emission towards SMS3 \citep{tieftrunk1998}. Our continuum source SMS3-MM2 is associated with the main NH$_3$ core, and MM1 and MM3 are associated with one of the compact NH$_3$ clumps, clump 4 \citep{tieftrunk1998}. From the line-width of the single dish $^{13}$CO$(2-1)$ data ($\sim$4.9~km~s$^{-1}$), following the equation described in \citet{pillai2006} we estimate the virial mass for SMS3 as $\sim1563$ $M_\odot$, which is much larger than the virial mass \citet{tieftrunk1998} obtained (300--400 $M_\odot$) from the NH$_3$ observations and the mass we {obtain} from the SCUBA measurements (Table \ref{conttable}). Our $^{13}$CO line may not be completely optically thin, and traces more or less all of the molecular gas, including the more diffuse envelope, whereas the NH$_3$ and SCUBA traces the dense gas in the center. Therefore, the NH$_3$ virial mass \citet{tieftrunk1998} obtained should better represent the mass from SCUBA. Hence, SMS3 is approximately in virial equilibrium. Since the peak column density (Table \ref{conttable}) is also above the proposed threshold for high-mass star formation of 1 g cm$^{-2}$ \citep{krumholz2008}, we infer that SMS3 is at a very early stage of massive star formation. 

For the more evolved regions SMS1 and SMS2, both harbor NIR clusters \citep{ojha2004, bik2011} and HCH{\scriptsize II} regions \citep{tieftrunk1997}, and both show strong signatures of active star forming activity. However, the continuum sub-sources in these two regions all show different evolutionary features.

\paragraph{Evolutionary sequence in SMS1} Our high resolution observations show a more complicated picture. Most of the molecular lines, especially some high excitation energy level sulfur-bearing molecular lines, peak on the strongest continuum source SMS1-MM1, which is associated with the HCH{\scriptsize II} region cluster W3 M and infrared source IRS5. \citet{rodon2008} resolved SMS1-MM1 into four 1.4 mm continuum sources. Combining the strong outflows we find here, we can confirm that a massive multiple system is forming in SMS1-MM1. The fact that SMS1-MM2 is not associated with any NIR point sources and has far fewer lines than SMS1-MM1 is consistent with a starless core. The high temperature of SMS1-MM2 we estimate (Table \ref{temp_table}) might be due to the shock heating caused by the outflows driven by SMS1-MM1 (Fig.~\ref{sms1_out}). SMS1-MM4 shows almost no lines aside from the three CO isotopologues, and it is also associated with an NH$_3$ core detected by \citet{tieftrunk1998}. Therefore we conclude that SMS1-MM4 might be at a very early evolutionary stage and a potential starless core. Since SMS1-MM4 is in the direction of the red-shifted component of outflow-b and there is SiO emission nearby, it might be a swept up ridge by the outflow.

\paragraph{Evolutionary sequence in SMS2} \citet{tieftrunk1998} found a filamentary structure traced in NH$_3$ starting at the HCH{\scriptsize II} region W3 Ca and extending towards the northeast. Our SMA observations not only recover part of the filament but also detect continuum emission towards the UCH{\scriptsize II} region W3 C, which is the most evolved source in SMS2. Most lines detected in SMS2 peak on the SMA continuum source SMS2-MM2, which is associated with the HCH{\scriptsize II} region W3 Ca \citep{tieftrunk1997}. We also detect outflows towards SMS2-MM2. The infrared source IRS4 detected by \citet{wynn-williams1972} consists of two infrared sources, which are IRS4-a and IRS4-b (Fig.~\ref{sms2_ks}). IRS4-a was classified as an O8-B0.5V star, and is considered to be the exciting source of UCH{\scriptsize II} region W3 C \citep{bik2011}. The other IR source IRS4-b is associated with the SMA continuum source SMS2-MM2 and the HCH{\scriptsize II} region, and it also dominates the mid- and far-infrared emission. All these features show that SMS2-MM2 is in the hot core evolutionary stage. It is surprising that the strongest continuum source in this region SMS2-MM1 shows only a few molecular lines and is in a younger evolutionary stage as a potential massive starless core. Last but not least, SMS2-MM3 and SMS2-MM7 exhibit fewer lines than SMS2-MM1 (Table \ref{spectable}) and have strong DCN emission (Fig.~\ref{sms2_others}), which makes them the youngest sources in SMS2. All these continuum sources show clear evolutionary stages from the most evolved UCH{\scriptsize II} region to potential starless cores within 30~000 AU. 

In summary, at large scales, the whole W3 Main region harbors the relatively evolved H{\scriptsize II} regions W3 A, W3 H and W 3 D (Fig.~\ref{cont}), and a young cluster is forming inside. Our observations reveal three massive star forming subregions in different evolutionary stages.  W3 SMS3 is in the youngest evolutionary stage, while W3 SMS1 and W3 SMS2 are more evolved. Furthermore, within the subregions, we find the SMA continuum sources in different evolutionary stages, from potential starless cores to typical high-mass hot cores. The data clearly show that the age difference of massive star formation regions not only exists on clump scales, but also within the clump.

\subsection{Astrochemistry}
Chemistry can be used to determine the evolutionary stages of massive star-forming regions. The continuum sources show different chemical evolutionary stages. SMS2-MM2 shows typical HMC chemical characteristics, with many hot/dense gas tracer emission, e.g. CH$_3$OH and CH$_3$CN lines (Fig.~\ref{sms2_others}). The $^{13}$CS and DCN emission offset from SMS2-MM2 also shows its HMC nature, since in a high temperature environment ($\sim$100 K) CS quickly reacts with OH forming SO and SO$_2$ \citep{beuther2009}, and DCN can be destroyed rapidly by reaction with atomic hydrogen \citep{wright1996, hatchell1998}. For the younger sources, i.e. the potential starless cores, DCN and $^{13}$CS lines are also useful to determine evolutionary stages.

DCN is reported as a indicator of the evolution of the dense cores, as it is liberated from the evaporated grain mantle and could be destroyed rapidly in the hot gas by reaction with atomic hydrogen \citep{wright1996, hatchell1998}. The ratio of DCN/HCN is often used to determine the evolutionary stage of high-mass cores. Since we do not have the HCN observations, we cannot get this ratio to constrain the exact evolutionary stages of the sources, however, the detection/non-detection of DCN provides us with a clue about the evolution. We detect DCN emission towards SMS2-MM1/3/7 and SMS3-MM2, which indicates that these four continuum sources are in an early evolutionary stage. The non-detection of DCN towards the starless cores SMS1-MM2 and SMS1-MM4 could be due to the fact that the jets driven by SMS1-MM1 heated the ambient gas and destroyed the DCN molecules. We do not detect DCN towards SMS3-MM1 and SMS3-MM3 either, but $^{13}$CS emission is detected towards these two sources, which indicates they are in very early evolutionary stage too.

SMS1-MM1 also shows HMC chemical characteristics, with many emission lines and all the SO, SO$_2$ isotopologues peaking on the continuum source (Figs. \ref{sms1_others} and \ref{sms1_so}). However, of the typical dense/warm gas tracers CH$_3$OH and CH$_3$CN, only very weak CH$_3$OH emission and no CH$_3$CN emission is detected towards the continuum peak (Fig.~\ref{sms1_others}). The offset peak of CH$_3$OH could be explained by the shock produced by the outflow, but we cannot explain why the CH$_3$CN peak is offset from the continuum peak. 

Another puzzle is that CH$_3$OH lines show extended emission towards the area between very young sources SMS3-MM1 and SMS3-MM2 (Fig.~\ref{sms3_others}). These are peculiar chemical properties of the SMS3 region.

\subsection{Sequential and potential triggered star formation}
NIR imaging and spectroscopy studies by \citet{bik2011} show an age spread of at least 2--3~Myr for the massive stars within the H{\scriptsize II} regions in W3 Main. X-ray studies by \citet{feiglson2008} revealed a low-mass young stellar cluster with an age of $>$0.5~Myr and \citet{ojha2004} obtained an age of the NIR cluster of 0.3$-$1 Myr form the $K_s$-band luminosity function. In contrast, the UCH{\scriptsize II} regions such as W3~C and W3~F only have an age of $\sim$10$^5$~yr \citep{wood1989,churchwell2002,mottram2011,davies2011}, and our SMA continuum sources SMS1-MM1 and SMS2-MM2 which are associated with HCH{\scriptsize II} regions are even younger (few times 10$^4$ yr, \citealt{mottram2011,davies2011}). Furthermore, the potential starless cores in SMS3 are still at the onset of massive star formation. The age differences reveal different stellar populations.  The OB stars associated with the diffuse H{\scriptsize II} regions and the low-mass cluster members are the first generation stars, and the exciting sources of the UCH{\scriptsize II} regions and our SMA continuum sources are the second generation stars. The second generation star formation might be triggered by the first generation stars or perhaps the star formation activities last over a prolonged period \citep{feiglson2008}.

\begin{figure*}[htbp]
   \centering
    \includegraphics[angle=0,width= 15cm]{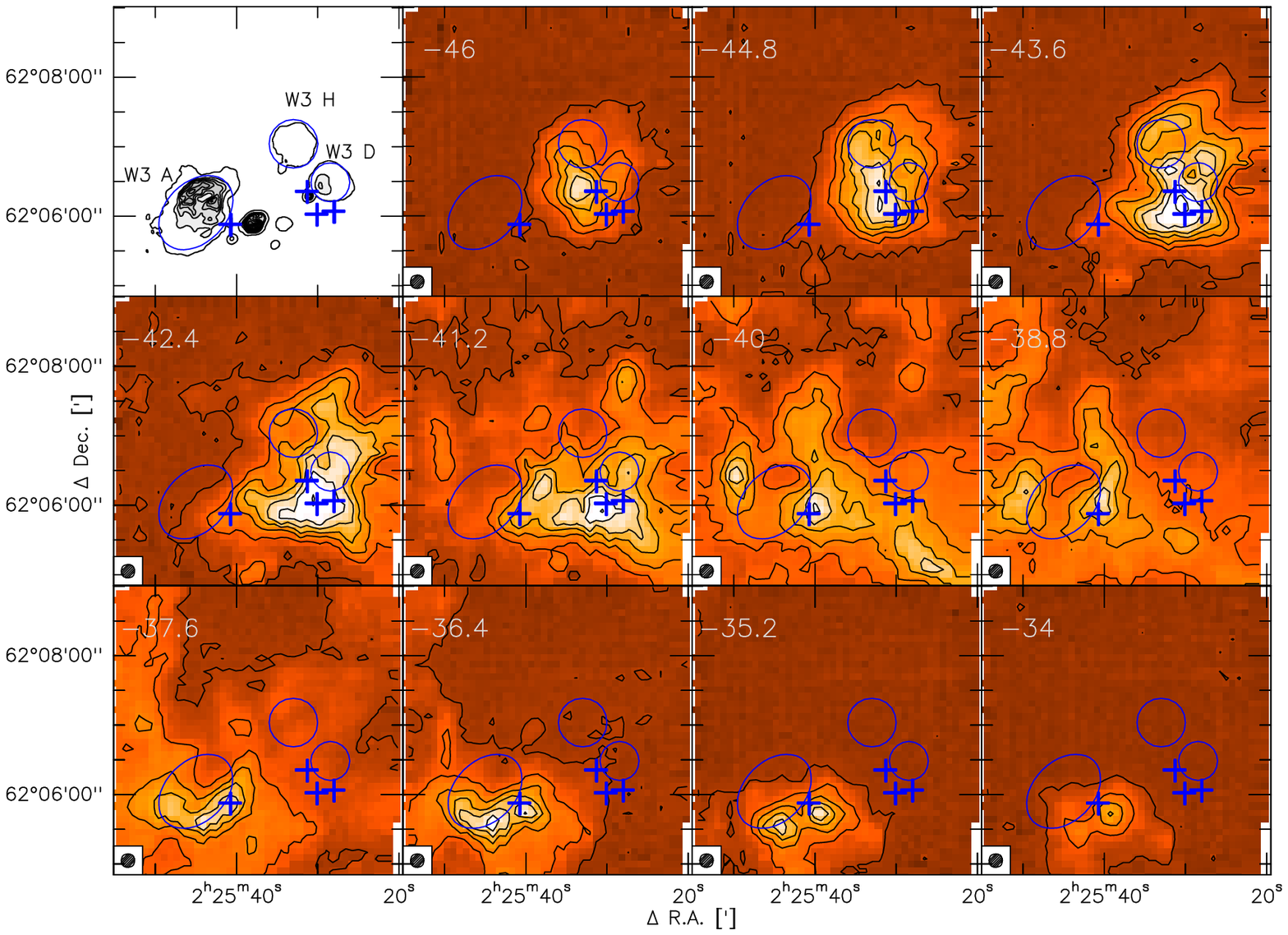}
    \caption{The single dish $^{13}$CO$(2-1)$ channel map and the 6~cm VLA continuum image \citep{tieftrunk1997} of W3 Main. The {\it top left panel} shows the 6 cm VLA continuum image, and the contour levels start at 2.8~mJy~beam$^{-1}$ and continue in steps of 50~mJy~beam$^{-1}$. The rest of the panels show the single dish $^{13}$CO($2-1$) channel map with a spectral resolution of 1.2~km~s$^{-1}$, with contour levels starting at 5$\sigma$ and continuing in steps of 20$\sigma$ (1$\sigma=0.35$~K). The crosses mark the position of the SMA continuum sources SMS1-MM1, SMS2-MM2, SMS3-MM1 and SMS3-MM2. The three circles/ellipses outline the approximate position and size of the three H{\scriptsize II} regions, W3~A, H, and D \citep{tieftrunk1997}. The beam of the single dish data is shown in the bottom left corner of each panel.}
\label{13co_chan}
\end{figure*}

To investigate the interactions between the H{\scriptsize II} regions and the molecular cloud, we construct the single dish $^{13}$CO($2-1$) channel map (Fig.~\ref{13co_chan}). The channel map shows that the molecular cloud forms circle-like structures with velocity gradients around the H{\scriptsize II} regions W3~A, H and D. For W3~A, in the panel $-40$~km~s$^{-1}$ of Fig.~\ref{13co_chan}, the molecular cloud forms a ``ring'' around W3~A with two main clumps located at the northeast and west of the H{\scriptsize II} region' s edge. These two main clumps move to the south a bit in the panel $-38.8$~km~s$^{-1}$, but the ``ring'' is still there. In the panel $-37$~km~s$^{-1}$, the ``ring'' is no longer present, and the clumps move to the south and join into one. Finally, in the panel $-34$~km~s$^{-1}$, the clump moves to the southwest. These structures indicate a velocity gradient of the molecular cloud around the H{\scriptsize II} region W3~A from the north to the south. A similar velocity gradient is also found around the H{\scriptsize II} region W3 H, which starts in the panel -$44.8$~km~s$^{-1}$ and the direction is from the east to the southwest. For the H{\scriptsize II} region W3~D, a ``ring'' structure is found in the panel $-43.6$~km~s$^{-1}$ of Fig.~\ref{13co_chan}, and the clumps move south and east of W3~D in the next three channels. These velocity gradients suggest that the H{\scriptsize II} regions are expanding and interacting with the ambient molecular gas. Furthermore, the YSOs and potential starless cores of the SMS1, SMS2, and SMS3 regions are located close to the edge of the H{\scriptsize II} regions. These facts together indicate that the H{\scriptsize II} regions may have even triggered the star formation in these sub-regions.

Similar structures with molecular gas distributed as a ring structure have also been found around many H{\scriptsize II} regions or bubbles \citep{deharveng2009,beaumont2010,fang2012}.

In the S255 complex, \citet{wang2011} revealed three massive star forming regions with different evolutionary stages within the same region. Based on the age difference between the low-mass cluster and the central high-mass protostellar objects, they suggest that the low-mass stars form first under pressure of the H{\scriptsize II} regions sitting around them, and then the low- to intermediate-mass stars may enhance the instability of the central high-mass cores and may potentially have triggered the formation of the central high-mass stars. Unlike the symmetric geometry of the S255 complex, W3 Main is more randomly distributed. 

SMS1 sits between W3~A and W3~B, and is also associated with the dense young stellar cluster \citep{feiglson2008, megeath1996}, which has the highest YSO density in W3 Main. The exciting source of W3~A has an age of $\sim$3 Myr \citep{bik2011}, and is much older than the NIR cluster (0.3--1~Myr, \citealt{ojha2004}). Therefore, W3~A may trigger the formation of the NIR cluster and even the central high-mass stars (IRS5). 

\citet{tieftrunk1998} found a filament along the edge of W3 H and W3 D from NH$_3$ observations. \citet{tieftrunk1998} suggested that this filament may be a swept-up layer of W3 H. As part of the filament, we suggest the SMA continuum sources in SMS2 may have formed under the influence of W3 D, and may even be triggered by W3 D in the dense gas at the head of the cometary H{\scriptsize II} region W3 D as discussed by \citet{hoare2007}. Additionally, we see an ``age gradient", from the southwest to the northeast, with the objects getting younger and younger (Fig.~\ref{cont}).

\section{Summary}
Combining the single dish IRAM 30~m observations and the high resolution SMA observations, we characterize the different evolutionary stages of massive star formation regions within the W3 Main complex. W3 SMS1, SMS2, and SMS3 show varying dynamical and chemical properties, which indicates that they are in different evolutionary stages. Even within each subregion, massive cores with different evolutionary stages are found, e.g. in SMS2, the SMA continuum sources show evolutionary stages from the most evolved UCH{\scriptsize II} region to potential starless cores within 30~000 AU. 

We find outflows in SMS1 and SMS2. The absence of an outflow in SMS3 indicates it is at the onset of massive star formation. The multiple outflows in SMS1 confirm a multiple system is forming. 

We detect 36 molecular lines in SMS1, 27 lines in SMS2 and 7 lines in SMS3. The temperature is determined from the CH$_3$CN $k$-ladder rotational diagram and H$_2$CO line intensity ratios for continuum sources in SMS1 and SMS2. 

Rotational structures are found for SMS1-MM1 and SMS2-MM2, and in SMS1-MM1 we find a velocity gradient in the hydrogen recombination line H30$\alpha$. This may trace rotation and/or outflow, but we cannot distinguish between these scenarios.

Evidence for potential interactions between the molecular cloud and the H{\scriptsize II} regions is found in the $^{13}$CO$(2-1)$ channel map, which may indicate triggered star formation.

\acknowledgments
Y. W. acknowledges the productive discussions with Min Fang, and the observers at the SMA and 30 m telescopes. 

This work is supported by the NSFC 10873037 and 10921063,10733030, China. 

We acknowledge the unknown referee.

{\it Facilities:} \facility{SMA}, \facility{IRAM:30m}
\end {CJK*}

\bibliographystyle{apj}
\bibliography{wang2012}{}

\clearpage
\begin{landscape} 
\begin{deluxetable*}{lrccccccccccccccccc}
\tabletypesize{\footnotesize}
\tablewidth{0pt}
\setlength{\tabcolsep}{0.1cm}
\tablecaption{Observed lines at all continuum peaks.}
\tablehead{
\colhead{$\nu$}&
\colhead{Transition}&
\colhead{{\it E}$_{\rm lower}$}&
\multicolumn{4}{c}{SMS1}&
\colhead{}&
\multicolumn{7}{c}{SMS2}&
\colhead{}&
\multicolumn{3}{c}{SMS3}\\
\cline{4-7}
\cline{9-15}
\cline{17-19}
\colhead{$[$GHz$]$}&
\colhead{}&
\colhead{$[$K$]$}&
\colhead{MM1}&
\colhead{MM2}&
\colhead{MM3}&
\colhead{MM4}&
\colhead{}&
\colhead{MM1}&
\colhead{MM2}&
\colhead{MM3}&
\colhead{MM4}&
\colhead{MM5}&
\colhead{MM6}&
\colhead{MM7}&
\colhead{}&
\colhead{MM1}&
\colhead{MM2}&
\colhead{MM3}
}
\startdata
\multicolumn{19}{c}{LSB}\\
\hline
\noalign{\smallskip}  
   217.105  &SiO($5-4$)			     &21	&$\surd$&$\surd$&~&$\surd$&~&~&~&~&~&~&~&~& ~&~&~&~\\
   217.239  &DCN($3-2$)				     &10	&~&~&~&~&~&$\surd$&~&$\surd$&~&~&~&$\surd$&	  ~&~&$\surd$&~\\  
   217.830  &$^{33}$SO($5_{6,6}-4_{5,5}$)	     &24	&$\surd$&$\surd$&~&~&~&~&~&~&~&~&~&~& ~&~&~&~\\ 
   218.222  &H$_2$CO($3_{0,3}-2_{0,2}$)		     &10	&$\surd$&$\surd$&~&~&~&$\surd$&$\surd$&$\surd$&~&~&~&$\surd$&	  ~&~&~&~\\ 
   218.325  &HC$_3$N($24-23$)			     &121	&$\surd$&~&~&~&~&$\surd$&$\surd$&~&~&~&~&~&	  ~&~&~&~\\ 
   218.440  &CH$_3$OH($4_{2,2}-3_{1,2}$)E	     &35	&~&$\surd$&~&~&~&$\surd$&$\surd$&~&~&~&~&~&	  ~&$\surd$&$\surd$&$\surd$\\	  
   218.476  &H$_2$CO($3_{2,2}-2_{2,1}$)		     &58	&$\surd$&$\surd$&~&~&~&$\surd$&$\surd$&$\surd$&~&~&~&~&	  ~&~&~&~\\ 
   218.760  &H$_2$CO($3_{2,1}-2_{2,0}$)		     &58	&$\surd$&$\surd$&~&~&~&$\surd$&$\surd$&~&~&~&~&~&	  ~&~&~&~\\ 
   218.903  &OCS($18-17$)			     &89	&$\surd$&$\surd$&~&~&~&~&$\surd$&~&~&~&~&~&  ~&~&~&~\\ 
   219.276  &SO$_2$($22_{7,15}-23_{6,16}$)	     &394	&$\surd$&$\surd$&~&~&~&~&~&~&~&~&~&~& ~&~&~&~\\ 
   219.355  &$^{34}$SO$_2$($11_{1,11}-10_{0,10}$)    &50	&$\surd$&$\surd$&~&~&~&~&~&~&~&~&~&~& ~&~&~&~\\ 
   219.466  &SO$_2$(v$_2$=1)($22_{2,20}-22_{1,21}$)     &983	&$\surd$&~&~&~&~&~&~&~&~&~&~&~& ~&~&~&~\\ 
   219.560  &C$^{18}$O(2$-$1)			     &5.3	&$\surd$&$\surd$&~&$\surd$&~&$\surd$&$\surd$&$\surd$&$-$&~&$-$&$\surd$&		  ~&$\surd$&$\surd$&$\surd$\\	  
   219.798  &HNCO(10$_{0,10}-$9$_{0,9}$)	     &48	&$\surd$&$\surd$&~&~&~&~&$\surd$&~&~&~&~&~&  ~&~&~&~\\ 
   219.949  &SO(6$_{5}-$5$_{4}$)		     &24	&$\surd$&$\surd$&~&$\surd$&~&$\surd$&$\surd$&$\surd$&~&~&~&$\surd$&	  ~&$\surd$&$\surd$&$\surd$\\	  
   220.079  &CH$_3$OH(8$_{0,8}-$7$_{1,6}$)E	     &85	&~&$\surd$&~&~&~&~&$\surd$&~&~&~&~&~&  ~&~&~&~\\ 
   220.165  &SO$_2$(v$_2$=1)($16_{3,14}-16_{2,14}$)  &613	&$\surd$&~&~&~&~&~&~&~&~&~&~&~& ~&~&~&~\\ 
   220.399  &$^{13}$CO(2$-$1)			     &5.3	&$\surd$&$\surd$&$\surd$&$\surd$&~&$\surd$&$\surd$&$\surd$&$-$&$\surd$&$\surd$&$\surd$&~&$\surd$&$\surd$&$\surd$\\	  
   220.585  &HNCO(10$_{1,9}-$9$_{1,8}$)	     	     &91	&$\surd$&~&~&~&~&~&$\surd$&~&~&~&~&~&  ~&~&~&~\\ 
   220.594  &CH$_3$CN(12$_6-$11$_6$)		     &315	&~&~&~&~&~&~&$\surd$&~&~&~&~&~&  ~&~&~&~\\ 
   220.620  &$^{33}$SO$_2$($11_{1,11}-10_{0,10}$)    &50	&$\surd$&~&~&~&~&~&~&~&~&~&~&~& ~&~&~&~\\ 
   220.641  &CH$_3$CN(12$_5-$11$_5$)		     &237       &~&~&~&~&~&~&$\surd$&~&~&~&~&~&  ~&~&~&~\\ 
   220.679  &CH$_3$CN(12$_4-$11$_4$)		     &173	&~&~&~&~&~&~&$\surd$&~&~&~&~&~&  ~&~&~&~\\ 
   220.709  &CH$_3$CN(12$_3-$11$_3$)		     &123	&~&$\surd$&~&~&~&$\surd$&$\surd$&~&~&~&~&~&  ~&~&~&~\\ 
   220.730  &CH$_3$CN(12$_2-$11$_2$)		     &87	&~&$\surd$&~&~&~&~&$\surd$&~&~&~&~&~&  ~&~&~&~\\ 
   220.743  &CH$_3$CN(12$_1-$11$_1$)		     &65	&~&$\surd$&~&~&~&~&$\surd$&~&~&~&~&~&  ~&~&~&~\\ 
   220.747  &CH$_3$CN(12$_0-$11$_0$)		     &58	&~&$\surd$&~&~&~&~&$\surd$&~&~&~&~&~&  ~&~&~&~\\ 
\hline
\noalign{\smallskip}
\multicolumn{19}{c}{USB}\\
\hline
\noalign{\smallskip}   
   229.348  &SO$_2$($11_{5,7}-12_{4,8}$)	     &111	&$\surd$&$\surd$&~&~&~&~&$\surd$&~&~&~&~&~&  ~&~&~&~\\ 
   229.545  &SO$_2$($13_{2,12}-13_{1,13}$)	     &828	&$\surd$&~&~&~&~&~&~&~&~&~&~&~& ~&~&~&~\\ 
   229.759  &CH$_3$OH(8$_{-1,8}-$7$_{0,7}$)E	     &77	&~&$\surd$&~&~&~&$\surd$&$\surd$&~&~&~&~&~&	  ~&~&~&~\\ 
   229.858  &$^{34}$SO$_2$(4$_{2,2}-$3$_{1,3}$)    &7.7	&$\surd$&$\surd$&~&~&~&~&~&~&~&~&~&~& ~&~&~&~\\ 
   230.027  &CH$_3$OH(3$_{2,2}-$4$_{1,4}$)E	     &28	&~&$\surd$&~&~&~&~&$\surd$&~&~&~&~&~&  ~&~&~&~\\ 
   230.538  &$^{12}$CO(2$-$1)			     &5.5	&$\surd$&$\surd$&$-$&$\surd$&	~&$\surd$&$\surd$&$\surd$&$\surd$&$\surd$&$\surd$&$\surd$&~&$-$&$\surd$&$\surd$\\	  
   230.965  &SO$_2$($37_{10,28}-38_{9,29}$)	     &881	&$\surd$&~&~&~&~&~&~&~&~&~&~&~& ~&~&~&~\\ 
   231.061  &OCS(19$-$18)			     &100	&$\surd$&~&~&~&~&~&$\surd$&~&~&~&~&~&  ~&~&~&~\\ 
   231.221  &$^{13}$CS(5$_0-4_0$)		     &22	&~&$\surd$&~&~&~&$\surd$&~&$\surd$&~&~&~&~&  ~&$\surd$&$\surd$&$\surd$\\	  
   231.901  &H30$\alpha$			     &~	 	&$\surd$&~&$\surd$&~&~&~&~&~&$\surd$&$\surd$&$\surd$&~& ~&~&~&~\\ 
   231.981  &SO$_2$($14_{3,11}-14_{2,12}$)	     &593	&$\surd$&~&~&~&~&~&~&~&~&~&~&~& ~&~&~&~\\ 
   232.266  &S$^{18}$O($5_6-4_5$)		     &37	&$\surd$&~&~&~&~&~&~&~&~&~&~&~& ~&~&~&~\\ 
   232.419  &CH$_3$OH($10_{2,8}-9_{3,7}$)A$+$	     &154	&$\surd$&$\surd$&~&~&~&~&$\surd$&~&~&~&~&~&  ~&~&~&~\\ 
   232.946  &CH$_3$OH($10_{-3,8}-11_{-2,10}$)E	     &179	&~&~&~&~&~&~&$\surd$&~&~&~&~&~&  ~&~&~&~
\enddata
\label{spectable}
\tablecomments{``$\surd$''indicates that this transition is detected towards this continuum source. ``--'' indicates that the spectrum extracted towards this source shows a pure negative feature due to the missing flux problem.}
\end{deluxetable*}
\clearpage
\end{landscape}

\end{document}